\documentclass[a4paper,fleqn]{cas-sc}

\UseRawInputEncoding
\usepackage[authoryear]{natbib}

\usepackage{float} 
\usepackage{subfigure}
\usepackage{algorithm}
\usepackage{algorithmic}
\usepackage{color}

\def\tsc#1{\csdef{#1}{\textsc{\lowercase{#1}}\xspace}}
\tsc{WGM}
\tsc{QE}
\tsc{EP}
\tsc{PMS}
\tsc{BEC}
\tsc{DE}

\begin{document}
\let\WriteBookmarks\relax
\def\floatpagepagefraction{1}
\def\textpagefraction{.001}

\shorttitle{TCAN for Popularity Prediction}

\shortauthors{X. Sun et~al.}

\title [mode = title]{Explicit Time Embedding Based Cascade Attention Network for Information Popularity Prediction}                      



%
\author[1]{Xigang Sun}[
]


\ead{xgsun27@stu.suda.edu.cn}


\credit{Conceptualization of this study, Methodology, Experiments, Writing - Original draft preparation}

\affiliation[1]{organization={Soochow University},
    city={Suzhou},
    postcode={215006}, 
    country={China}}

\author[1]{Jingya Zhou}[]
\cormark[1]
\ead{jy_zhou@suda.edu.cn}
\credit{Funding acquisition, Conceptualization of this study, Supervision, Writing - review \& editing}

\author[2]{Ling Liu}[]
\ead{lingliu@cc.gatech.edu}

\credit{Writing - review \& editing}

\author[2]{Wenqi Wei}[]
\ead{wenqiwei@gatech.edu}

\credit{Writing - review \& editing}

\affiliation[2]{organization={Georgia Institute of Technology},
    city={Atlanta},
    postcode={30332}, 
    country={USA}}







\cortext[cor1]{Corresponding author}



\begin{abstract}
Predicting information cascade popularity is a fundamental problem in social networks. Capturing temporal attributes and cascade role information (e.g., cascade graphs and cascade sequences) is necessary for understanding the information cascade. Current methods rarely focus on unifying this information for popularity predictions, which prevents them from effectively modeling the full properties of cascades to achieve satisfactory prediction performances. In this paper, we propose an explicit Time embedding based Cascade Attention Network (TCAN) as a novel popularity prediction architecture for large-scale information networks. TCAN integrates temporal attributes (i.e., periodicity, linearity, and non-linear scaling) into node features via a general time embedding approach (TE), and then employs a cascade graph attention encoder (CGAT) and a cascade sequence attention encoder (CSAT) to fully learn the representation of cascade graphs and cascade sequences. We use two real-world datasets (i.e., Weibo and APS) with tens of thousands of cascade samples to validate our methods. Experimental results show that TCAN obtains mean logarithm squared errors of 2.007 and 1.201 and running times of 1.76 hours and 0.15 hours on both datasets, respectively. Furthermore, TCAN outperforms other representative baselines by 10.4\%, 3.8\%, and 10.4\% in terms of MSLE, MAE, and R-squared on average while maintaining good interpretability.
\end{abstract}


\maketitle

\section{Introduction}

\begin{figure*}[t]
\centerline{\includegraphics[width=0.6\textwidth]{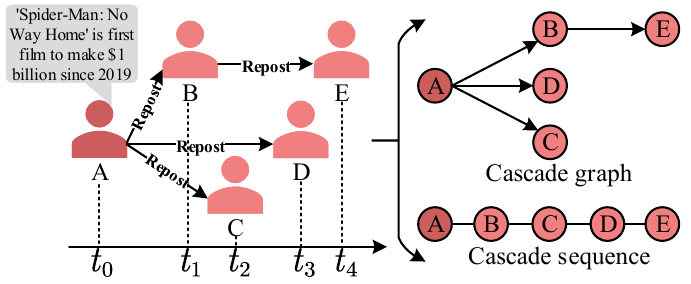}}
\caption{An example of an information diffusion cascade. \textbf{Left}: The cascade is a directed acyclic temporal graph (DATG), where the root user A posts a piece of content at $t_0$ and others repost it successively. \textbf{Right}: The cascade graph and sequence are extracted from the DATG.}
\label{fig1}
\end{figure*}

The prevalence of online social networks has been profoundly changing our daily life. Social media users often share interesting content (e.g., literals~\citep{FanLLZ21}, images~\citep{WangWCHMZ20}, videos~\citep{XieZZPYHLC20}, etc.) with one another.  As shown in Figure~\ref{fig1}, the information is usually diffused in a cascading manner~\citep{islam2018deepdiffuse} on social platforms, 
e.g., users join in the reposts of a specific post on Weibo or Twitter. Thus the information popularity can be measured by the number of involved users~\citep{CaoSGWC20}. Many merchants make decisions based on the information popularity for gaining profits, such as advertisements~\citep{guanfeng2020tkde,XinLZLZBZ21} and recommendations~\citep{YalcinB21,DBLP:journals/tweb/WuLCG20}. In addition, public opinion disasters such as misinformation about the COVID-19 will burden the virus prevention and control, and cyber violence will bring negative effects on people, while most of these catastrophes can be significantly alleviated or even avoided by predicting cascades in advance~\citep{bian2020rumor, ChenZZB21}. We can also quantify the impact of the publications, as well as individual scholars/authors by analyzing citation cascades contained in scientific databases~\citep{XuZLTZ22}. 

Cascade prediction research usually falls into two levels of prediction tasks: micro (next user prediction~\citep{SankarZK020}) and macro (popularity prediction). We pay more attention to the latter in this paper,
and we aim to exploit the observed information cascade process to predict its future popularity and apply this task to broad domains. Recently, many studies~\citep{ fan2021a,GaoCLYCT19} have focused on the task of information popularity prediction, including regression (e.g., cascade user size~\citep{li2017deepcas} and repost volume~\citep{cao2017deephawkes} prediction) and classification (e.g., outbreak prediction~\citep{CuiJYWZY13}) problems, and they can further be classified into three main categories: Feature-extraction methods~\citep{ZhaoWCLW18,cheng2014can} are committed to identifying and extracting useful features inside cascades, including content features, temporal features, structural features, and user behavior features, and then these features are fed into classical machine learning models to make predictions. These methods primarily rely on domain knowledge and lack sufficient generalization capabilities. Generative models ~\citep{shen2014modeling, zhao2015seismic} are designed based on time-series data, they build probability statistical models by utilizing point processes (e.g., Poisson Process and Hawkes process) for modeling the process of repost arrival. They are inherently interpretable, but usually fail to provide accurate predictions. Deep learning based models ~\citep{gao2020deep} predict information popularity by employing recurrent neural networks (RNNs) on sampled diffusion sequences or graph neural networks (GNNs) on cascade graph episodes. We classify these methods without exploiting real time (or explicit) features as implicit time learning. The end-to-end learning schema endows them with powerful predictability and generalization capability, but they still suffer from several big challenges:

\begin{enumerate}
\itemsep=0pt
\item \textbf{Fail to consider node temporal attributes.} 
Real time feature refers to the repost time point of the cascade node, which reflects the temporality of cascades. As shown in Figure~\ref{fig1}, each node joins in the reposting process at different time points. We transform real time features into intrinsic temporal attributes, i.e., periodicity, linearity, and non-linear scaling. Injecting the temporal attributes into nodes can help models capture the fine-grained temporal relationship between nodes and learn the law of popularity changes over time. But few work embeds temporal attributes into node features. 
 
\item \textbf{Incomplete cascade role information learning.} 
Cascade roles are vital to predicting the information popularity, e.g., cascade graphs and cascade sequences reflect the diffusion structures and the process of node joining. However, current works omit either the structure information or the joining information of nodes, and cannot fully capture the cascade role information.

\item \textbf{Cumbersome cascade processing.} 
Existing methods try to exploit sequence sampling or episode division to model the temporality of cascades. However, the former cannot capture the complete graph structure information, while the latter lacks a unified standard for division, and both manners are cumbersome for cascade processing.
\end{enumerate}

We therefore propose a novel popularity prediction architecture, explicit \textbf{T}ime embedding based \textbf{C}ascade \textbf{A}ttention \textbf{N}etwork \textbf{(TCAN)} to address the above challenges. The motivation of our TCAN is to capture accurate temporal relationships and comprehensive cascade role information. Our main contributions are summarized as follows:

\begin{enumerate}
\itemsep=0pt
\item \textbf{General explicit time embedding approach.} 
To tackle the temporality of cascades, we propose a general explicit cascade time embedding (TE) approach to endow nodes with time-awareness, which leverages time functions to convert temporal attributes into time vectors and embeds them directly into the cascade node representations. Compared with the sequence sampling and episode division methods, TE is explicit and simple with no cumbersome pre-processing required before time embedding. Moreover, it can be easily generalized to other domains.

\item \textbf{Cascade role learning architecture.} 
To explore the cascade role information comprehensively, we design a novel cascade role learning architecture, which contains a cascade graph attention encoder (CGAT) and a cascade sequence attention encoder (CSAT). The only premise of role learning is injecting temporal attributes into node features by TE. CGAT can effectively capture the diffusion structure by performing local attention on cascade graphs. CSAT employs the Bidirectional Long Short Term Memory (Bi-LSTM) with an attention pooling for capturing the contextual features of node joining in cascade sequences. We can also capture the fine-grained temporal relationship between cascade nodes by employing the CGAT and CSAT.

\item \textbf{Extensive experiment evaluations.} We conduct extensive experiments on real-world datasets, demonstrating that our model outperforms the leading-edge methods. The enhanced experiments show that our TE not only improves the prediction performance significantly but also has a good generalization capability. We also conduct explanation experiments to show that the cascade representations learned by TCAN are highly effective for popularity prediction, which shows the good interpretability of TCAN.
\end{enumerate}

The rest of the paper is organized as follows. Section~\ref{sec2} reviews related work. Section~\ref{sec3} states research preliminaries. We describe our proposed TCAN in detail in Section~\ref{sec4}. Some important experiments are presented in Section~\ref{sec5}. Finally, Section~\ref{sec6} summarizes our work.

\section{Related Work}
\label{sec2}
We review a series of existing works regarding information diffusion popularity prediction from two aspects of traditional methods and deep learning methods.

\subsection{Traditional Methods}

The traditional methods of popularity prediction are divided into two categories: feature-extraction methods and generative models.

The former extracts a series of hand-crafted features and uses classical machine learning models to make predictions. Content-related features such as domain URLs and topic features~\citep{martin2016exploring}, hashtags and mentions~\citep{SuhHPC10}, and sentiments~\citep{NaveedGKA11} are extracted from source post content to assist popularity prediction.
Structural features~\citep{KupavskiiOUUSGK12,gao2014effective,WangYHYLW16} (e.g., friendship networks, repost networks, and centrality) and temporal features~\citep{odriguez2014uncovering,mishra2016feature,shulman2016predictability} (e.g., repost time, time interval, and accumulated repost count) are considered by researchers to be more effective for predicting popularity. More works~\citep{cheng2014can,tsur2012what's,hong2011predicting} combine the above features and user-related information (e.g., user profiles or activity statistics) for improving prediction accuracy. Furthermore, the performance of machine learning models heavily depends on the quality of extracted features. These methods mainly rely on the specific features chosen with experience and cannot be handled with a unified standard, which prevents them from being generalized across different domains.

The latter focuses on modeling the conditional intensity function (i.e., point process) for the popularity accumulation of each content and employs maximum likelihood to infer model parameters. Poisson Process (PP) is a kind of point process where the intensity function is constant, and models based on PP predict popularity dynamics via incorporating PP into the Bayesian architecture. For example, models~\citep{shen2014modeling, gao2015modeling} introduce the reinforcement mechanism into PP, called RPP, for characterizing the "richer-get-richer" phenomenon. MPP~\citep{rong2015why} mixes three cascading influence Poisson processes to make predictions. Another point process is Hawkes Process (HP) which describes self-exciting properties. Literatures~\citep{cao2017deephawkes, zhao2015seismic, bao2015modeling, KobayashiL16} use the Self-Excited Hawkes Process to characterize the process of individual content gaining popular, and model each repost's contribution, user influence, and time decay effect. Furthermore, in~\citep{mishra2016feature}, a feature-based predictive layer is built on top of the Self-Excited Hawkes Process to enhance the model performance. Generative models offer good interpretability for the popularity dynamics in information diffusion by using solid mathematical theory. However, they are unable to make full use of the useful information hidden in cascades for popularity prediction.

\subsection{Deep Learning Based Models}

Deep learning based models are designed to automatically learn the
representations of input cascades through backpropagation. Some existing graph representation learning methods~\citep{zhou_csur} can be employed to embed cascade graphs. DeepWalk~\citep{perozzi2014deepwalk}, Node2vec~\citep{grover2016node2vec}, and LINE~\citep{tang2015line} sample paths for graph embedding by random walk strategies, but the random manner is easy to miss some structural information. GNN-based methods, such as GCN~\citep{kipf2017semi-supervised}, Graphsage~\citep{hamilton2017inductive} and GAT~\citep{velickovic2018graph}, can effectively learn the graph structure information through neighbor aggregation. However, they are unable to make popularity predictions directly due to the limitations of the cascade temporality. GRU~\citep{chung2014empirical} and LSTM~\citep{hochreiter1997long,DBLP:conf/bigdataconf/WenZYSSY19} can learn the cascade temporality information by modeling the information diffusion sequence.

Therefore, current deep learning models designed for popularity prediction mainly combine graph representation learning~\citep{DBLP:conf/cdmake/SchlottererSG17,DBLP:journals/toit/HuangLGLAC21} and sequence learning~\citep{chung2014empirical} together. Some representative methods are reviewed as follows: 
DeepCas~\citep{li2017deepcas} samples node repost sequence via random walks on cascade graphs and combines all the sequence representations for prediction. 
DeepHawkes~\citep{cao2017deephawkes} learns the embedding of each repost path and integrates Hawkes Process into a deep learning architecture at the same time, and finally it pools all path representations by weighting the time decay parameters for prediction. Topo-LSTM~\citep{wang2017topolstm}, and TempCas~\citep{TangLHXZS21} are similar works that fully exploit diffused paths in the cascade.
CasCN~\citep{chen2019information} employs CasLaplacian for each sub-cascade episode that is separated from a cascade graph in chronological order, and then uses LSTM to learn episode sequences for prediction. Similar to CasCN, GSAN~\citep{Huang00ZZ20} utilizes graph attention networks and Transformers~\citep{vaswani2017attention} to model the context features of each sub-cascade episode. CasSeqGCN~\citep{WANG2022117693} uses GCN based on dynamic routing aggregation and LSTM to model cascade snapshot sequences for prediction. VaCas~\citep{0002XZTZ20} and MuCas~\citep{ChenZZB22} are similar work based on cascade episode division. 
These implicit time learning methods lack the guidance of accurate temporal attributes and omit some cascade role information, which prevents them from accurately capturing the pattern of the popularity increment over time. Inspired by previous work on temporal feature learning~\citep{kazemi2019time2vec,xu2020inductive,wang2020joint}, we propose the explicit time embedding and cascade role learning to avoid the shortcomings of implicit time learning and help TCAN to achieve superior prediction performances.

\section{Preliminaries}
\label{sec3}
In this section, we make some necessary explanations for the cascade and the popularity prediction task.
Let us take post $P_i$ as an example, its diffusion process during the observation time forms a cascade $C_i$. We extract two types of information from the cascade in accordance with the cascade roles, i.e., cascade graph and cascade sequence, as shown in Figure~\ref{fig1}.

\textbf{Definition 1: Cascade.} The process of information $P_i$'s diffusion forms a cascade, which is defined as a directed acyclic temporal graph (DATG) $C_i=(V_i, E_i, T_i)$, for $v_{i,a},\ v_{i,b}\in V_i$, and $t_{i,a}, \ t_{i,b} \in T_i$, if $v_{i,b}$ reposts $P_i$ from $v_{i,a}$ at time $t_{i,b}$, also known as $v_{i,b}$ is influenced by $v_{i,a}$ to join $C_i$, and then there has a directed temporal edge $(v_{i,a},v_{i,b},t_{i,b}) \in E_i$. All the above nodes, edges, and repost time points form the cascade.

\textbf{Definition 2: Cascade Graph.} A cascade graph $G_i=(V_i, E_i)$ released from $C_i$ is a directed acyclic graph (DAG) without repost time information. It retains all the nodes and non-temporal edge relationships in $C_i$.

\textbf{Definition 3: Cascade Sequence.} We detach all nodes which join in $C_i$ formed by $P_i$, and sort them according to the repost time to obtain a cascade sequence $S_i$. $S_i=\{v_{i,0},v_{i,1},...,v_{i,n-1}\}$ is a node sequence, and $n$ is the length of cascade sequence $S_i$, i.e., $n=|S_i|$.

The post's popularity is measured by the number of involved users in the information diffusion process during a specified time window. In this paper, we focus on the incremental popularity of posts from the observation time to the end time. Hence, we convert the popularity prediction to a regression task defined below:

\textbf{Definition 4: Popularity Prediction.} Given the observation time window $[0,t_b]$ and the end time $t_e$, the observed popularity size of $P_i$ is $L^{t_b}_i$. The objective is to learn a regression function $f(\cdot)$, for every $\{G_i, S_i, T_i\}$ from $C_i$, to predict the incremental popularity $\Delta L_i=L_i^{t_e}-L_i^{t_b}$, formalized as $\{G_i, S_i, T_i\} \stackrel{f(\cdot)}{\longrightarrow} \Delta L_i$.

\section{TCAN}
\label{sec4}
\begin{figure*}[t]
\centerline{\includegraphics[width=1\textwidth]{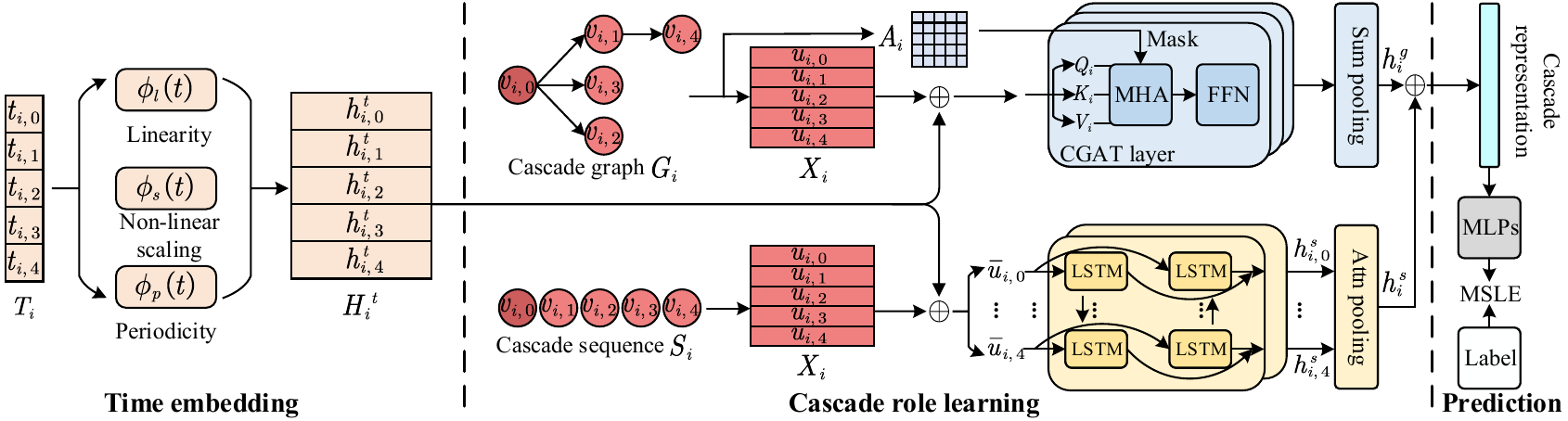}}
\caption{The overview of TCAN architecture, where $\bigoplus$ denotes concatenation operation. CGAT and CSAT correspond to the top and bottom of the cascade role learning, respectively.}
\label{fig2}
\end{figure*}

In this section, we introduce the architecture of TCAN in detail. As shown in Figure~\ref{fig2}, it mainly contains three components: 
1) Time embedding (TE) integrates the temporal attributes into initial node features. It learns a continuous time mapping function that maps the value of each time point to a $d_t$\textit{-dimensional} vector, and then concatenates the time vectors with initial node features. 2) In the cascade role learning part, we adopt the adjacency matrix mask attention based on a scaled dot-product to build the CGAT layer for cascade graph embedding. The cascade sequence is inputted into Bi-LSTM layers with attention pooling to obtain the sequence embedding. We concatenate the cascade graph and sequence embedding to generate cascade representations.    
3) Prediction layer integrates cascade representations into MLPs to make predictions.

\subsection{Explicit Time Embedding}
Time information reflects not only the position of each influenced node but also the repost time interval between nodes in the cascade. Integrating temporal attributes into node features endows nodes with time-awareness, which can help the model capture the fine-grained temporal relationship between cascade nodes and further uncover the law of information diffusion. For example, popularity may accumulate more easily when an item is reposted in a short time, while the popularity will decay if the repost time interval is long. Therefore, we design a general explicit time embedding (TE) approach to achieve the fusion of time and node features.

Our TE aims to map the input time sequence $T_i$ into a time vector matrix. We first describe how TE converts generalized times into vectors, and then use $T_i$ as the input of TE to output a time embedding matrix. We define the influence temporal relationship between nodes as the dot-product of node's time vectors:
\begin{equation}
R_{(t_1, t_2)}:=<\phi(t_1), \phi(t_2)>,
\end{equation}
where $\phi(\cdot)$ is a continuous time mapping function and $R$ denotes the temporal relationship. How to design the time mapping function is the key point. As we know, time has the attributes of both periodic change and linear progress. For instance, we start counting from 0 after 24 hours every day while time passes linearly. In addition, some potential non-linear factors also need to be considered. These attributes are also  closely related to people's publishing behaviors and popularity growth processes, and we give a detailed analysis in Section~\ref{sete}. In order to capture these temporal attributes, the time vectors generated by $\phi(\cdot)$ should possess the periodicity, linearity, and non-linear scaling. Thus, we define the mapping functions of these attributes, respectively.

\textit{(1) Periodicity.}
We employ the cosine or sine function to capture the periodicity of time. Let us take cosine as an example, the function is defined as follows:
\begin{equation}
\phi_p(t)=cos(\omega_p*t+b_p),
\end{equation}
where $\omega_p$ and $b_p$ are the frequency and phase-shift of the cosine function, and will be updated by our model training. The cosine function's period is $\frac{2 \pi}{\omega_p}$, so we get the same value for $t$ and $t+\frac{2\pi}{\omega_p}$. If $\omega_p=\frac{2\pi}{24}$, cosine function can capture the changes of behavior with a period of 24 hours. We can obtain the time embedding $h_{tp}\in \mathbb{R}^{1 \times d'_t}$ with periodicity from the trained cosine function where $d'_t$ is the embedding dimension, and $d'_t \geq 1$ endows time with multiple potential periods. 

\textit{(2) Linearity.}
Linear transformation is one of the most basic operations, which represents the linear progress of time. We define the linear transformation of time as follows:
\begin{equation}
\phi_l(t)=w_l*t+b_l,
\end{equation}
where $w_l \in \mathbb{R}^{1\times 1}$ and $b_l\in \mathbb{R}^{1\times 1}$ are weight and bias parameters, and both of them are learnable. We obtain the time embedding $h_{tl} \in \mathbb{R}^{1\times 1}$ with linearity after training.

\textit{(3) Non-linear scaling.}
{\color{black}
We also need to capture some time-dependent non-linear factors related to the popularity growth. For example, the popularity growth trend is often observed to decay over time and the repost timestamps show a power-law distribution~\citep{gao2015modeling}. Therefore, we introduce a non-linear mapping function $\phi_s(\cdot)$ to describe it:}
\begin{equation}\label{eq4}
\phi_s(t)=w_s*\sqrt{t},
\end{equation}
where the sqrt operation scales the range of timestamps, and $w_s \in \mathbb{R}^{1\times1}$ is a learnable scaling parameter. The non-linear result we obtained is denoted by $h_{ts} \in \mathbb{R}^{1\times1}$. Here we can select logarithmic or exponential functions to replace Eq.~\ref{eq4}, but they might induce gradient vanishing and explosion problems and are not within the scope of this paper.

It is worth noting that we only learn the weight parameters of each time attribute function to adapt to different time units of each dataset, rather than directly fitting all cases with MLPs. This learnable temporal attribute selection method reduces noise as much as possible while generating efficient low-dimensional temporal embeddings.
We ultimately obtain the embedding vector $h_t \in \mathbb{R}^{1 \times d_t}$ of time $t$ via the concatenation:
\begin{equation}
h_t = concatenation\{h_{tp},h_{tl},h_{ts}\},
\end{equation}
where the dimension of time embeddings is $d_t=d'_t+2$. For time vector $h_t$, the first $d'_t$ dimensions contain periodic information, and the latter two dimensions contain linear and non-linear information, respectively.

Information popularity is often related to the participating nodes' own features (i.e., initial node features)~\citep{cheng2014can}, which can be used to learn cascade representations. We use $X_i=\{u_{i,0},u_{i,1},...,u_{i,n-1}\} \in \mathbb{R}^{n \times d}$ to denote the initial node features, and $d$ is the feature dimension. In our deep learning framework, we directly learn the node's feature representation from historical cascade data to adapt to specific scenarios following~\citep{cao2017deephawkes}. Specifically, we initialize a node feature matrix $F \in \mathbb{R}^{M \times d}$ that obeys the standard normal distribution, which is shared by all users, where $M$ represents the total number of users in all cascades. For each node $v_{i,j}$ in the cascade $C_i$, there is a corresponding one-hot vector $e_{v_{i,j}} \in \mathbb{R}^{1\times M}$. Then the initial vector representation of $v_{i,j}$ can be obtained by using $e_{v_{i,j}}$ to look up in $F$, i.e., $u_{i,j}=e_{v_{i,j}}F$. The node feature matrix $F$ will be supervised by the incremental  popularity prediction target and learned during TCAN training.

The time embedding $H_i^t \in \mathbb{R}^{n \times d_t}$ is generated by employing TE on $T_i$. We combine the node's own features $X_i$ and the time embedding matrix $H_i^t$ to generate new node features $\bar{X}_i$ by concatenation, i.e., $\bar{X}_i=concatenation\{X_i, H_i^t\}$, and $\bar{X}_i=\{\bar{u}_{i,0},\bar{u}_{i,1},...,\bar{u}_{i,n-1}\} \in  \mathbb{R}^{n \times (d+d_t)}$, which is the input of the subsequent CGAT and CSAT modules.

    
\subsection{Cascade Graph Embedding}
Cascade graph is a critical cascade role that reflects the information diffusion structures and the influence relationships between nodes. For example, nodes with similar structures have similar popularity contributions~\citep{lilian2013virality}, and central nodes are more influential than leaf nodes and are more inclined to contribute to the popularity. Besides, diffusion graph structures also reflect spatial bursts (depth and breadth) that show a rapid diffusion of information during the observation time. To fully capture the cascade structural information and the cascade temporal relationship, we propose the cascade graph attention encoder (CGAT) which improves the Transformer encoder~\citep{vaswani2017attention} by incorporating graph structure information to encode the cascade graph $G_i=(V_i, E_i)$. The cascade graph attention layer (CGAT layer) is the core of the encoder, which performs a local attention on $G_i$. The local attention adopts the self-attention mechanism with an adjacency matrix mask, where the scaled dot-product corresponds to our temporal relationship function $R$.

\begin{figure}[t]
\centerline{\includegraphics[width=0.25\textwidth]{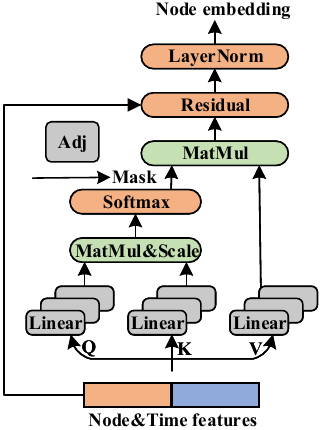}}
\caption{One head cascade graph attention layer. It is extended to a multi-head attention mechanism  (MHA) and used in the CGAT layer box in Figure~\ref{fig2}. \textit{MatMul\&Scale} represents the scaled dot-product and \textit{Adj} refers to the adjacency matrix.}
\label{fig3}
\end{figure}

Let $A_i$ denote an directed adjacency matrix of $G_i$. Suppose $Q_i$, $K_i$, and $V_i$ be the queries, keys, and values that are generated from the cubic linear transformations of the input node features $\bar{X}_i$, i.e.,
\begin{equation}\label{eq6}
Q_i = \bar{X}_iW_Q, \quad K_i=\bar{X_i}W_K, \quad V_i=\bar{X}_iW_V,    
\end{equation}
where $W_Q,\ W_K,\ W_V \in \mathbb{R}^{(d+d_t)\times (d+d_t)}$ are weight parameters that are used to extract the features of time-awareness nodes. Then the self-attention scores $\alpha_i \in \mathbb{R}^{n \times n}$ of cascade graph nodes $V_i$ are calculated by masked scaled dot-product of queries and keys with a softmax function, i.e.,
\begin{equation}
\alpha_i = softmax(mask_{A_i}(\frac{Q_i K_i^T}{\sqrt{d+d_t}})),
\end{equation}
where $mask_{A_i}(\cdot)$ is the adjacency matrix mask function that is employed to prevent mixing non-adjacent relations into self-attention scores while acting as graph structure information, and $\sqrt{d+d_t}$ is a scaled term. 
We then compute the output hidden representation as:
\begin{equation} \label{eq8}
H_i^g=\alpha_iV_i,
\end{equation}
and we use $Attn(\cdot)$ to denote the calculation process of Eq.~\ref{eq6}-\ref{eq8}. Compared with GAT~\citep{velickovic2018graph}, CGAT uses the self-attention mechanism to support parallel computing, and each node in the cascade graph receives the features from its neighbor nodes and aggregates them according to the importance of influence and temporal relationships. We define a CGAT layer as:
\begin{equation}
\begin{split}
H_i^g &= Attn(\bar{X}_i, A_i, [W_Q,W_K,W_V]),   \\
\bar{H_i^g} &= LN(FFN(H_i^g)+\bar{X}_i),    \\
\end{split}
\end{equation}
where $FFN$ is the feed forward network block, and then the Residual~\citep{he2016deep} and LayerNorm~\citep{ba2016layer} (LN) are used to stabilize parameters and prevent gradient explosion.  Figure~\ref{fig3} illustrates the process of one head cascade graph attention layer.

Furthermore, in order to improve performances and stabilize training for the CGAT layer, we extend the attention to multi-head attention (MHA). Let $k$ be the number of heads, and we obtain the outputs of self-attention from a total of $k$ different heads, i.e.:
\begin{equation}
\begin{split}
H_i^g&=concatenation\{H_{i,0}^g,...,H_{i,k-1}^g\}W_O, \\
\bar{H_i^g} &= LN(FFN(H_i^g)+\bar{X}_i), \\
\end{split}
\end{equation}
where $W_O\in \mathbb{R}^{k(d+d_t)\times(d+d_t)}$ is a projection parameter matrix. CGAT aggregates the high-order neighbor information of nodes by stacking multiple layers, and the $m$-th CGAT layer is calculated as:
\begin{equation}
\bar{H_i^g}^{(m)} = CGATLayer^{(m)}(\bar{H_i^g}^{(m-1)} ,A_i).
\end{equation}

Finally, we need to select a graph readout function to aggregate the node representations from the last CGAT layer into a cascade graph representation $h_i^g$. Sum pooling (SP) can make graph representations more discriminative and accurate, and also can suppress noise to reduce model overfitting to a certain extent~\citep{XuHLJ19}. Obviously, we focus on the whole cascade graph representation, so we use the SP to achieve the final graph readout process.

\subsection{Cascade Sequence Embedding}
Cascade sequence is another important cascade role that is relevant to the process of nodes joining cascade and accordingly is also closely related to the cascade size within the observation time. For example, previous nodes involved in a cascade may cumulatively influence the reposting behavior of subsequent nodes, namely the sequential temporal memory effect~\citep{WangSLGC17}, and the scale of future popularity is intrinsically correlated to the size of the observed cascade. The node joining process corresponds to the information diffusion sequentially over time (i.e., temporal diffusion), which encourages us to design the cascade sequence attention encoder (CSAT) to capture the features of influenced nodes and aggregate them into the cascade sequence representation.

To capture such cascade role information,
we utilize recurrent neural networks (RNNs) for learning the representation of the temporal diffusion sequence $S_i=\{v_{i,0},v_{i,1},...,v_{i,n-1}\}$, and its feature matrix is $\bar{X_i}$. However, the traditional RNN is not sensitive to long-term dependence due to the vanishing gradient problem.
To this end, the long short-term memory (LSTM) is used to alleviate such problem by adding a gating memory mechanism. The LSTM cell is formulated as follows:
\begin{equation}
    \begin{split}
        j_{i,t} &= \sigma(\bar{u}_{i,t}W_{xj}+h_{i,t-1}W_{hj}+b_j), \\
        f_{i,t} &= \sigma(\bar{u}_{i,t}W_{xf}+h_{i,t-1}W_{hf}+b_f), \\
        o_{i,t} &= \sigma(\bar{u}_{i,t}W_{xo}+h_{i,t-1}W_{ho}+b_o), \\
        \tilde{c_{i,t}}&=tanh(\bar{u}_{i,t}W_{xc}+h_{i,t-1}W_{hc}+b_c), \\
        c_{i,t} &= f_{i,t} \odot c_{t-1} + j_{i,t} \odot \tilde{c}_{i,t}, \\
        h_{i,t} &= o_{i,t} \odot tanh(c_{i,t}),
    \end{split}
\end{equation}
where $W_{x*}, W_{h*}$ and $b_*$ are weight and bias parameters, $j_{i,t}$, $f_{i,t}$, $o_{i,t}$, and $\tilde{c}_{i,t}$ denote the input gate, forget gate, output gate, and candidate memory cell, respectively.

We use the LSTM to learn cascade sequence node features from left to right, which controls the inflow of information through memory cells and gating mechanisms to simulate the temporal diffusion process. But we still need to learn sequence information from right to left to let early nodes know which nodes are influenced~\citep{li2017deepcas}. Furthermore, since nodes are embedded with time information, the forward LSTM allows historical time information to flow into the current node and enables the node to perceive future time nodes; the backward LSTM allows previous nodes to know at which time node the information diffusion behavior will occur in the future. Therefore, in order to obtain rich hidden information with diffusion possibilities, CSAT employs two bi-directional LSTM (Bi-LSTM) layers to model the diffusion cascade sequence and capture the temporal contextualized information of node joining behaviors. Accordingly, we obtain the node-level representation ${H_i^s}^{(1)}$: 
\begin{equation}
\begin{split}
\overrightarrow{H_i^s}^{(0)} &= LSTM_{fwd}(\bar{X}_i), \overleftarrow{H_i^s}^{(0)}=LSTM_{bwd}(\bar{X}_i),\ {H_i^s}^{(0)}=concatenation\{\overrightarrow{H_i^s}^{(0)}, \ \overleftarrow{H_i^s}^{(0)}\}, \\
\overrightarrow{H_i^s}^{(1)} &= LSTM_{fwd}({H_i^s}^{(0)}), \  \overleftarrow{H_i^s}^{(1)}=LSTM_{bwd}({H_i^s}^{(0)}), \ {H_i^s}^{(1)}=concatenation\{\overrightarrow{H_i^s}^{(1)}, \ \overleftarrow{H_i^s}^{(1)}\}, \\
lhs_i^s &= concatenation\{LHS(\overrightarrow{H_i^s}^{(1)}),LHS(\overleftarrow{H_i^s}^{(1)})\},
\end{split}
\end{equation}
where $LHS(\cdot)$ means to take the last hidden state of the output. We define the output of the last layer as ${H_i^s}:={H_i^s}^{(1)}, {H_i^s}\in\mathbb{R}^{n \times d_h}$, where $d_h$ is the hidden dimension of Bi-LSTM.

We usually use the last hidden state $lhs_i^s$ of the Bi-LSTM to represent the information of the entire sequence. But such an approach is not fully suitable for cascade popularity prediction. Even with Bi-LSTM we cannot obtain the full historical information of the cascade, and the prediction performance may degrade as the cascade size increases. To tackle above challenge, we employ self-attention pooling (Attn pooling, AP) on all node representations that are generated by the last Bi-LSTM layer. AP can help us to capture globally important temporal diffusion information by using the attention mechanism to identify key hidden states. The cascade sequence-level representation is calculated as:
\begin{equation}\label{eq14}
    h^s_i = AP(H_i^s),
\end{equation}
where AP consists of self-attention~\citep{vaswani2017attention} and sum pooling. To this way, the cascade sequence representation $h_i^s \in \mathbb{R}^{1 \times d_h}$ is readily used for cascade popularity prediction.

\begin{algorithm}[t]
\renewcommand{\algorithmicrequire}{\textbf{Input:}}
\renewcommand{\algorithmicensure}{\textbf{Output:}}
\caption{Batch learning with TCAN}
\label{alg1}
\begin{algorithmic}[1]
\REQUIRE Time sequence $T=\{T_0,T_1,...,T_{N-1}\}$, set of cascade graphs $G=\{G_0,G_1,...,G_{N-1}\}$, and cascade sequence $S=\{S_0,S_1,...,S_{N-1}\}$ from an entire training dataset.
\ENSURE The parameters $\Theta=\{\Theta_T,\Theta_G,\Theta_S\}$ of TCAN and the predicted incremental popularity $\hat{y}=\{\hat{y_0},\hat{y_1},...,\hat{y_{N-1}}\}$.
\STATE Initialize learnable parameters $\Theta=\{\Theta_T,\Theta_G,\Theta_S\}$ and node features $X=\{X_0,X_1,...,X_{N-1}\}$;
\STATE Obtain $H^t$ by applying TE to $T$;
\STATE Combine $H^t$ and $X$ together to obtain a new $\bar{X}$;
\WHILE {not convergence and training steps less than a predefined value}
\FORALL {$batch=\{G_{batch}, S_{batch}, \bar{X}_{batch}\}$ in $G, S, \bar{X}$}
\STATE Obtain the cascade graph representation $H^g_{batch}$ via calling the CGAT on $G_{batch}$, $\bar{X}_{batch}$;
\STATE Obtain the cascade sequence representation $H^s_{batch}$ via calling the CSAT on $S_{batch}$, $\bar{X}_{batch}$;
\STATE Combine the $H^g_{batch}$ and $H^s_{batch}$ into MLP to get $\hat{y}_{batch}$;
\STATE Minimize the MSLE loss function $\mathcal{L}(y_{batch}, \hat{y}_{batch})$ to update the parameters $\Theta=\{\Theta_T,\Theta_G,\Theta_S\}$ by means of back propagation.
\ENDFOR
\ENDWHILE
\STATE Return the trained parameters $\Theta=\{\Theta_T,\Theta_G,\Theta_S\}$ and the predicted incremental popularity $\hat{y}=\{\hat{y_0},\hat{y_1},...,\hat{y_{N-1}}\}$.
\end{algorithmic}
\end{algorithm}

\subsection{Prediction and Computational Complexity Analysis}
Eventually, we obtain the cascade graph level-representation $h_i^g$ from CGAT and the cascade sequence level-representation $h_i^s$ from CSAT, where both vectors represent the temporal features of cascade graph and cascade sequence, respectively. Then, we concatenate and feed them into MLPs to get the predicted incremental popularity of $C_i$:
\begin{equation}
\begin{split}
h_i^{gs} &= concatenation\{h_i^g,h_i^s\}, \\
\hat{y}_i &= MLPs(h_i^{gs}).  
\end{split}
\end{equation}

Given the true incremental popularity $y_i$, our training loss is defined as the Mean Square Logarithmic Error (MSLE), i.e.,
\begin{equation}
\mathcal{L}(y_i, \hat{y}_i) = \frac{1}{N}\sum_{i=0}^{N-1}(log_2y_i-log_2\hat{y_i})^2,    
\end{equation}
where $N$ is the number of cascades in the training dataset. Algorithm \ref{alg1} describes the batch learning process of our TCAN, where $\Theta=\{\Theta_T, \Theta_G, \Theta_S\}$ denotes the set of learnable parameters of TE, CGAT, and CSAT in TCAN.

Let $n$ be the number of cascade nodes, $d_t$, $d$, and $d_h$ be the dimension of time vectors in TE, initial node features, and latent vectors in CSAT, respectively. TE module can be simply viewed as a $d_t$\textit{-dimensional} mapping process, so the complexity of TE is $\mathcal{O}(n\times d_t)$. The computational complexity of CGAT is mainly dominated by self-attention mechanism which is $\mathcal{O}(n^2\times d)$~\citep{vaswani2017attention}. Our CSAT is implemented based on LSTM, so its complexity is $\mathcal{O}(n\times d\times d_h)$. Therefore, the total computational complexity of TCAN is $\mathcal{O}(n\times d_t+n^2\times d+n\times d\times d_h)$.

\begin{table}[pos=b]
    \caption{The cascade information statistics of Weibo and APS.}
    \begin{tabular*}{\tblwidth}{@{} LLL @{}}
    \toprule
    Dataset &Weibo (before|after filtering) &APS (before|after filtering) \\
    \midrule
    Nodes&6,738,040|4,045,652 &616,316|418,057 \\
    Reposts/Cites&15,311,973|7,742,959 &3,304,400|1,107,012 \\
    Number of cascades (posts/papers) &119,313|34,935 &616,316|27,800 \\
    \midrule
    Avg. cascade size &174.02|298.57&14.03|63.69\\
    Avg. path length &2.34|2.37 &4.13|5.43 \\
    Avg. repost interval &343.44|202.19 (seconds) &320.19|149.38 (days) \\
    \midrule
    After filtering & & \\
    Training samples &24,455 &19,460\\
    Validation samples &5,240 &4,170\\
    Test samples &5,240 &4,170 \\
    \bottomrule
    \end{tabular*}
    \label{t1}
    \end{table}

\section{Experiment Evaluation}
\label{sec5}
In this section, we first introduce some details about real-world datasets and benchmark models, and then show the results of our comparative experiments, ablation studies, enhancement experiments, efficiency analysis together with interpretability studies.

\begin{figure}[b]
\centering
\subfigure[Weibo]{
\includegraphics[width=1\textwidth]{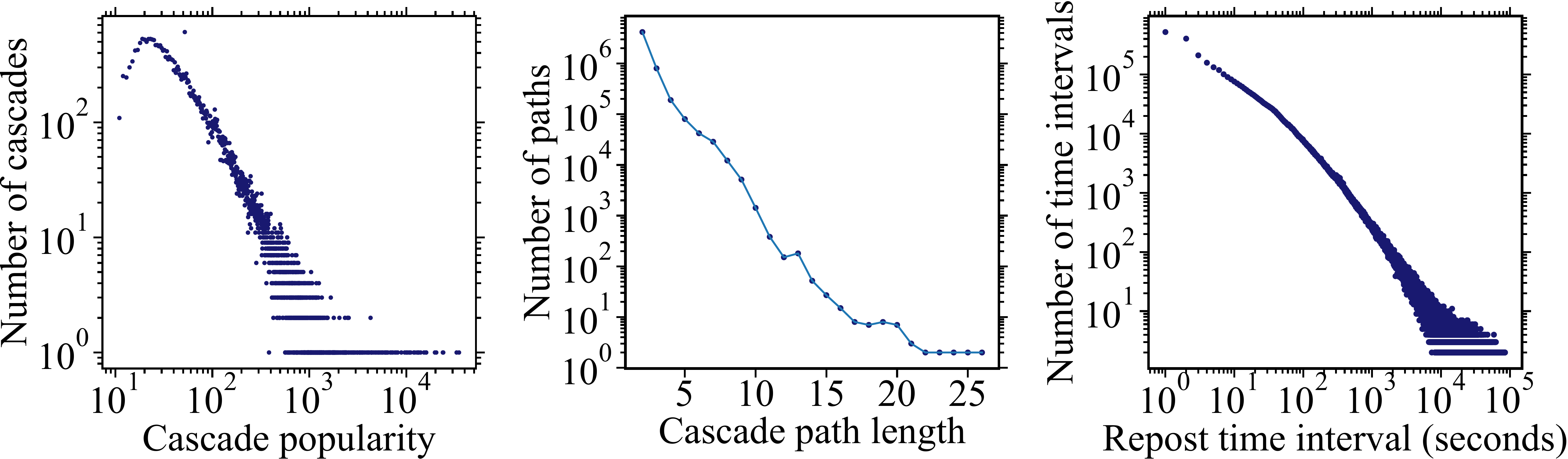}
}
\subfigure[APS]{
\includegraphics[width=1\textwidth]{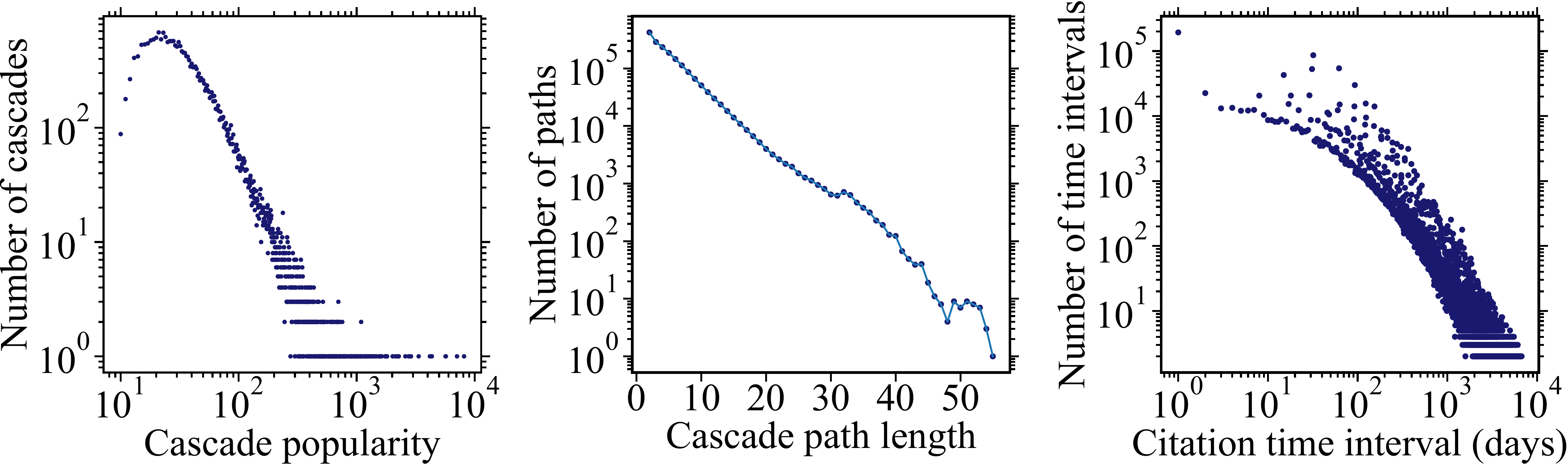}
}
\caption{Distributions of cascade popularity, path length, and repost time interval on filtered Weibo (up) and APS (down).}
\label{fig4}
\end{figure}

\subsection{Datasets}
In our experiments, we use two kinds of public datasets, namely the media dataset Weibo~\citep{cao2017deephawkes} and the scientific dataset APS~\citep{shen2014modeling}, and both are widely used in current studies.  

\textit{Sina Weibo.} The Weibo dataset is collected from a popular micro-blog platform in China, where hundreds of millions of users actively join in the diffusion of information every day. This dataset intercepts the posts released on June 1, 2016, and tracks all reposts within the next 24 hours. Each repost cascade consists of the repost paths of a post and the corresponding timestamps, where the unit of repost time is set as seconds. For example, user $A$ releases a post, and user $C$ repost the post by user $B$ at time $t_C$, we record the repost path $A/B/C:t_C$.

\textit{American Physical Society.} The APS dataset contains lots of scientific papers published on American Physical Society (APS) journals from 1983 to 2017. Each piece of data is a cascade that reflects the topology of the paper's citations. Specially, when a target paper $A$ appears in paper $B$'s reference list published at $t_B$, we record the citation path $A/B:t_B$, and the unit of citation time is day. For each paper, if it cites the target paper and other new papers that also cite the target paper as a source, we only keep the 
latest citation path.

We further filter the datasets according to specific rules and then use them to train our model. For the Weibo dataset, following~\citep{cao2017deephawkes,chen2019information}, we conventionally process the dataset by retaining only the posts released between 8:00 am and 6:00 pm. 
For each cascade, the observation time window is set as 1 hour by default, and the prediction time is set as 24 hours, i.e., the total users who join in the reposting process within 24 hours are the final popularity. After filtering out the cascades whose observation size is less than 10, all the cascade samples are divided into a training set (24,455), a test set (5,240), and a validation set (5,240) according to the ratio of 7:1.5:1.5.
For the APS dataset, we focus on papers published before 1997 and guarantee a minimum 20-year citation period for each paper, i.e., the number of papers within 20 years is the final popularity. Cascades with an observed cascade size of less than 10 are also filtered out. We set the default observation time as 3 years and split all citation cascades into a training set (19,460), a test set (4,170), and a validation set (4,170) based on the same ratio.

Table \ref{t1} shows the statistics before and after filtering on two datasets, each post/paper corresponds a cascade sample. Through data filtering, we retain 29\% and 5\% of the useful cascade samples on two datasets with 60\% and 68\% of the number of nodes and 51\% and 34\% of the reposts/cites of the original datasets. 44\% (52,154) and 88\% (543,593) of small cascades (the observed sizes are less than 10) on two datasets are filtered out, and their average cascade size is about 25. No matter whether filtering operation is performed or not, the cascade size of Weibo is much larger than that of APS, however, the average repost path length of Weibo is shorter than that of APS. It indicates that predicting the popularity of social media content are harder than predicting on scientific data.
On the filtered sub-datasets, both the average cascade size and the average path length become higher than that on the original datasets, and the average repost time decreases due to discarding meaningless small cascades. Data filtering enables our model to focus on the popularity prediction of relatively large cascades, which makes more sense in real-world application scenarios. We also add a variant of TCAN, TCAN-Origin, running on original datasets to test the impact of the filtered operation.
We further show the distribution of cascade popularity, path length, and repost time interval within filtered datasets in Figure \ref{fig4}. In both datasets, the distribution of cascade popularity follows a power-law distribution. That is, low-popularity cascades are abundant, while high-popularity cascades are rare. This phenomenon also appears in the repost time interval distribution. This means that there are few repost behaviors with long time intervals. The distribution of path lengths follows an exponential distribution, consistent with results reported in~\citep{cao2017deephawkes}.

    
\subsection{Baselines}
We select the following nine representative popularity prediction baselines for comparison, including a generative method (TiDeH), two feature-extraction methods (Feature-linear and Gradient Boosting Decision Tree (GBDT)), and six deep learning based methods. Furthermore, we regard the last six baselines as implicit temporal learning methods, which are further subdivided into two categories: sequence sampling (DeepCas, DeepHawkes, and Topo-LSTM) and episode division (CasCN, GSAN, and CasSeqGCN). The above baselines are described in detail as follows:

\begin{itemize}
\item \textit{Feature-linear, GBDT}: We follow literature~\citep{cheng2014can} to select 7 types of hand-crafted features that include both temporal and structural features: mean time interval between each repost, the observed cascade size, the cumulative popularity (every 10 minutes for Weibo, every 3 months for APS), the number of leaf nodes, average node degree, average and max length of repost path. We feed these features to a linear regression model with L2-regularization to make predictions, and we call this model Feature-linear. We also select an effective ensemble learning algorithm, Gradient Boosting Decision Tree (GBDT), as another feature-extraction baseline, and we train the GBDT with the above cascade features to predict popularity.

\item \textit{TiDeH}~\citep{KobayashiL16}: A generative baseline for popularity prediction. For each cascade, it builds a Time-Dependent Hawkes process model to imitate the popularity acquisition process. The window-based observed repost time and structural information are exploited as model inputs to estimate parameters.

\item \textit{DeepCas}~\citep{li2017deepcas}: It is the first end-to-end deep learning model designed for popularity prediction. DeepCas samples paths via a heuristic random walk, and utilizes the Bi-GRU with attention mechanism to generate cascade graph representations and then further predicts the incremental popularity.

\item \textit{DeepHawkes}~\citep{cao2017deephawkes}: It integrates three factors of Hawkes self-exciting point, i.e. influence of users, self-exciting mechanism, and time decay effect, into a deep learning architecture to make predictions, so as to enhance its interpretability. It can be viewed as a combination of a generative model and a deep learning model.

\item \textit{Topo-LSTM}~\citep{wang2017topolstm}: It uses a novel topological recurrent neural network for modeling dynamic DAGs and endows node representations with rich structural features. Topo-LSTM is designed to predict the next influenced user rather than the popularity, and we modify its classification layer to adapt to our regression task.

\item \textit{CasCN}~\citep{chen2019information}: It splits the cascade graph into episodes, and employs cascade graph convolution on each episode to generate a sub-cascade level representation, and then utilizes LSTM to capture the episode sequence information.

\item \textit{GSAN}~\citep{Huang00ZZ20}: GSAN is similar to CasCN, it uses the graph attention network and Transformer instead of cascade graph convolution and LSTM to capture the dynamic context information of each episode.

\item \textit{CasSeqGCN}~\citep{WANG2022117693}: CasSeqGCN is also similar to CasCN, it sets the step size for dividing the cascade snapshots to 5 by default. A GCN based on dynamic route aggregation is employed to learn cascade snapshot representations, and then an LSTM is used to capture snapshot dynamics.

\item \textit{TCAN-RNN}: It is a variant of TCAN. In CSAT, we only use the last layer hidden state $lhs_i^s$ of Bi-LSTM instead of AP as the cascade sequence representation.

\item \textit{TCAN-Original}: We train and test TCAN on the original datasets (i.e., before filtering). 

\item \textit{TCAN-OriFil}: We train TCAN on original datasets and test on the filtered test datasets.

\end{itemize}

\subsection{Parameter Settings}
\emph{Traditional baselines (Feature-linear, GBDT, TiDeH):} We employ Ridge regression in python's scikit-learn library, and the regularization strength alpha is chosen from \{0.01,0.02,0.03,...,1\}; For GBDT, the number of estimators is chosen from \{20,40,60,...,200\}; For TiDeH, we adjust the observation and prediction period for adapting our datasets. \emph{Deep learning based baselines:} For the general hyper-parameters, the learning rates of the model and user embedding are chosen from \{1e-4,1e-3,1e-2,...,1\}; the L2 regularization coefficient is chosen from \{1e-3, 1e-2, 0.1, 1\}; the dropout rate is chosen from \{0.1,0.2,0.3,...,0.6\}. The embedding dimensionality of users is set to 50, and the input batch size is set to 32. For the specific hyper-parameters of the model, the sequence batch size and the sequence length in DeepHawkes and TopoLSTM are consistent with DeepCas; the cascade episode step in CasCN and GSAN is fixed to 1, and the CasSeqGCN's episode step is chosen from \{1,2,3,4,5\}. We pick the optimal hyper-parameters for the above chosen in the range to obtain the best validated MSLE.

\emph{TCAN:} We implement our TCAN by using PyTorch-Lightning~\citep{falcon2019pytorch}. The dimension of both time embeddings and node features is set to 32. We adopt three CGATLayers, and its hidden dimension is 64. CSAT adopts a two-layer Bi-LSTM and its hidden size is 64. For the TCAN output, we adopt three-layer MLPs for prediction, and the hidden dimensions are 64, 64, 1. Adam~\citep{kingma2015adam} is used to optimize model parameters during TCAN training, where the learning rate is 1e-3. We set the weight decay for Adam as 5e-4 and set the dropout rates for CGAT layers and MLPs as 0.1, respectively, to prevent model overfitting.
For all the deep learning models, we adopt early stopping for obtaining the best prediction results when the validation loss has not declined for 10 consecutive epochs.

\begin{table}[b]
    \caption{Performance comparison with baselines on Weibo (1 hour) and APS (3 years), and prediction time windows are 24 hours and 20 years, respectively. The best results are marked with bold and '-' denotes negative value.}
    \begin{tabular*}{\tblwidth}{@{} LLLLLLL @{}}
    \toprule
    Dataset &\multicolumn{3}{C}{Weibo} &\multicolumn{3}{C}{APS}\\
    \midrule
    Metrics& MSLE & MAE & $R^2$ & MSLE & MAE& $R^2$ \\
    \midrule
    Feature-linear&3.141 &1.321 &0.384&1.469 &0.949 &0.371 \\
    GBDT&2.934 &1.284 &0.425&1.456 &0.941 &0.377 \\
    TiDeH&8.724 &2.523 &-&6.825 &2.162 &- \\
    \midrule
    DeepCas&3.159 &1.281 &0.374&1.909 &1.073 &0.183 \\
    DeepHawkes&2.605 &1.133 &0.500&1.306 &0.888 &0.440 \\
    Topo-LSTM&2.788 &1.194 &0.447&1.549 &0.965 &0.336 \\
    \midrule
    CasCN&2.433 &1.123 &0.415&1.590 &0.984 &0.318 \\
    GSAN&2.295 &1.062 &0.539&1.455 &0.945 &0.377 \\
    CasSeqGCN&3.271 &1.294 &0.360&1.878 &1.064 &0.196 \\
    \midrule
    TCAN-RNN &2.352 &1.064 &0.539 &1.216 &0.854 &0.479 \\
    TCAN-Original& {\textbf{1.982}}& 1.044& 0.502& 1.274& 0.890& {\textbf{0.543}} \\
    TCAN-OriFil& 2.298& 1.094& 0.550& 2.237& 1.199& 0.042 \\
    \textbf{TCAN}&2.007 &\textbf{1.032} &\textbf{0.607}&\textbf{1.201} &\textbf{0.845} &0.486 \\
    \bottomrule
    \end{tabular*}
    \label{t2}
    \end{table}
    
Furthermore, we adopt three metrics to evaluate our approach, including Mean Square Logarithmic Error (MSLE), Mean Absolute Error (MAE), and R-squared ($R^2$). They are defined as follows:
\begin{equation}
\begin{split}
&MSLE = \frac{1}{N}\sum_{i=1}^N(log_2\hat{y_i}-log_2y_i)^2,\\
&MAE = \frac{\sum_{i=1}^{N}|log_2\hat{y_i}-log_2y_i|}{N}, \\
&R^2 = 1 - \frac{\sum_{i=1}^N(log_2\hat{y_i}-log_2y_i)^2}{\sum_{i=1}^N(log_2 y_i-\bar{log_2 y})^2},
\end{split}
\end{equation}
where $y_i$ is the target value, and  $\hat{y_i}$ is the predicted value, $\bar{log_2 y}$ is the average of the target value after logarithm. MSLE is the main evaluation metric and is widely used in previous work~\citep{li2017deepcas,cao2017deephawkes,Huang00ZZ20}. Due to the power-law distribution of cascade popularity, the model often favors small-label cascade samples and may make less accurate predictions for large-label cascade samples. MSLE will amplify errors when the model makes inaccurate predictions. Thus, MSLE can measure the predictive ability of the model for large-label cascades. MAE reflects the overall true prediction error of the model. It can measure the predictive ability of the model for small-label cascade samples. For the first two metrics, their range is $[0,+\infty)$, lacking an accurate upper bound. $R^2$ is used to judge the fitting effect of the regression model, and its upper limit is 1. We treat the model with lower MSLE and MAE values (approaching to 0) and higher $R^2$ value (approaching to 1) as the better baseline for the popularity prediction task.

All experiments are based on the same preprocessed datasets for popularity prediction. The hardware facilities are four RTX 2080Ti GPUs with 4*12 GB video memory.
    
\subsection{Comparison with Baselines} \label{cwb}
The overall performance comparison results are shown in Table \ref{t2}. TCAN beats all the baselines based on three metrics on both datasets. Specifically, TCAN exceeds the best baseline (GSAN) by 12.5$\%$, 2.8$\%$, and 11.2$\%$ on MSLE, MAE and, $R^2$ on Weibo, and exceeds the best baseline (DeepHawkes) by 8.0$\%$, 4.8$\%$, and 9.5$\%$ on three metrics on APS. The results demonstrate the superiority of TCAN against baselines. We can also conclude from Table~\ref{t2} as follows:

\textbf{(1)} For evaluation metrics: Since the popularity follows a power-law distribution (as shown in Figure~\ref{fig4}), and the observation time is much smaller than the prediction time, these situations make the task of predicting incremental popularity more difficult than ordinary regression tasks. So the $R^2$ of all methods is not high, but TCAN still maintains the highest performance on all datasets. In addition, the improvement of TCAN on the MSLE is higher than that of MAE, indicating that TCAN can more accurately predict large-label cascades.

\textbf{(2)}
For data filtering operations, we analyze the impact of filtered operations on TCAN from the following two perspectives: 1) We train and test TCAN on the original datasets (denoted by TCAN-Original). MSLE and MAE do not change obviously (average changes of 1.2\% and 5.4\% on Weibo and APS) compared to $R^2$ after filtering on both datasets. $R^2$ is improved by 17.3\% on Weibo, because meaningless small cascades only account for 44\% on original datasets, and relatively large cascades impact the value of $R^2$. The increase in the total proportion of these relatively large cascades after filtering contributes to a significant improvement on $R^2$. In comparison, the presence of a large number of meaningless small cascades on APS (account for 88\%) dominates the value of $R^2$, and filtering out these cascades leads to a significant reduction (about 10.5\%) on $R^2$. 2) We also record the results of TCAN trained on the original datasets and tested on the filtered test datasets (denoted by TCAN-OriFil). 
The results indicate that if we train TCAN on the original datasets, it will reduce the model's ability to generalize to relatively large future cascades.


\textbf{(3)} For traditional methods: TiDeH has strong assumptions on the settings of parameter and observation time, and does not mine latent cascade information, resulting in the worst prediction performance. Feature-extraction methods, i.e., Feature-linear and GBDT, are not always worse than deep learning models when perform on both datasets. On media Weibo dataset, Feature-extraction methods outperform DeepCas and CasSeqGCN on both metrics of MSLE and $R^2$,  but perform less effective compared with other methods. On scientific APS dataset, Feature-extraction methods outperform DeepCas, Topo-LSTM, CasCN, and CasSeqGCN on three metrics and perform relatively well. The reason is twofold: 1) Although hand-crafted features positively contribute to popularity prediction, it is difficult to generalize to different domains. 2) Some deep learning based methods fail to capture both cascade role information and explicit time information, making them worse than Feature-extraction methods.

\textbf{(4)} For sequence sampling based methods: DeepCas performs the worst on both datasets, because it simply uses RNNs to learn the sequences sampled from cascade graphs. Topo-LSTM does not have good performance due to the same reason. DeepHawkes performs better than other sequence sampling methods (DeepCas and Topo-LSTM), especially on APS, where it stands out from all baselines. The result benefits from the modeling of all diffused sequences, which contributes to capturing the law of diffused processes. 

\textbf{(5)} For episode division based methods: CasSeqGCN has strict requirements for both the episode division step size and the observation time, which makes it less versatile for datasets and ineffective for popularity prediction based on short observation time. GSAN considers the time information of each episode when using Transformer, which makes it superior to CasCN. However, GSAN cannot model the temporal relationship between cascade nodes and ignores the fine-grained node joining process. TCAN can capture the temporal relationship of nodes, and takes full advantage of both cascade structure and sequence information at the same time, so it outperforms GSAN and other baselines. 

\textbf{(6)} For TCAN-RNN: TCAN-RNN only uses the last hidden state $lhs_i^s$ of the Bi-LSTM for popularity prediction and fails to capture the complete cascade sequence information. As a result, it performs worse than TCAN, which in turn verifies the effectiveness of AP (Eq.~\ref{eq14}).

\subsection{Enhancement Study} \label{ehs}
We introduce three vanilla models (i.e., DeepWalk, GCN, and RNN) for enhancement experiments to verify that our proposed TE is general and effective. 

\begin{table}[pos=t]
\caption{Enhancement study about TE on Weibo (1 hour) and APS (3 years). TempDW, TempGCN, and TempRNN correspond to DeepWalk, GCN and RNN with TE, respectively, and 'Temp**-PL' denotes the naive model enhanced by TE without the non-linear scaling term.}
\begin{tabular*}{\tblwidth}{@{} LLLLLLL @{}}
\toprule
Dataset &\multicolumn{3}{C}{Weibo} &\multicolumn{3}{C}{APS}\\
\midrule
Metrics &MSLE &MAE &$R^2$ &MSLE &MAE &$R^2$\\
\midrule
DeepWalk&4.139 &1.535 &0.189 &2.081 &1.129 &0.109\\
GCN&3.486 &1.422 &0.312 &1.788 &1.046 &0.234\\
RNN&2.873 &1.221 &0.437 &1.407 &0.922 &0.397\\
\midrule
TempDW-PL&3.936 &1.532 &0.228 &2.047&1.128&0.123\\
TempGCN-PL&3.229&1.348 &0.367 &1.626&1.000&0.303\\
TempRNN-PL&2.555 &1.184&0.499&1.395&0.918&0.403\\
\midrule
TempDW &3.859&1.504 &0.244 &1.981&1.119 &0.152\\
TempGCN &3.135 &1.342 &0.385 &1.535 &0.966 &0.342\\
TempRNN &2.406 &1.138 & 0.528 &1.384 &0.917 &0.407\\
\bottomrule
\end{tabular*}
\label{t3}
\end{table}

\begin{itemize}
\item  \textit{DeepWalk~\citep{perozzi2014deepwalk}}: Performing random walk on a graph to generate node sequences, and then use the word2vec technique to learn the node or graph representations. We obtain the pre-trained cascade node features by employing DeepWalk on cascade graphs, and feed them into MLPs to make predictions.

\item  \textit{GCN~\citep{kipf2017semi-supervised}}: A scalable semi-supervised graph representation learning method by performing convolution operations on graphs. We apply GCN to extract the cascade graph features and predict the incremental popularity by MLPs.

\item  \textit{RNN~\citep{schuster1997bidirectional}}: In contrast to GCN, we substitute the GCN parts with Bi-LSTM to generate cascade sequence representations. The others remain unchanged.

\item \textit{TempDW}, \textit{TempGCN}, and \textit{TempRNN} are enhanced versions of three vanilla models after adding TE module. Adding '-PL' means that the TE module keeps only periodic and linear function terms.
\end{itemize}

Table \ref{t3} shows the results of the enhanced experiment, where the last three rows of the table report the experimental results of the enhanced vanilla models. We draw the following conclusions:

\textbf{(1)} The performance of the vanilla models can be greatly improved by adding a TE module. It proves that integrating temporal attributes into nodes by TE is conducive to accurately predicting popularity, and our TE module is effective and general. Furthermore, removing the non-linear scaling term (i.e., Temp**-PL) causes the predictive performance of all naive models to drop, verifying that the addition of the non-linear scaling in TE is meaningful.

\textbf{(2)} On both datasets, GCN (or TempGCN) is superior to DeepWalk (or TempDW) based on all metrics. It is mainly because GCN can capture sufficient structure information of cascade graphs. 

\textbf{(3)} Whether time enhancement is adopted or not, RNN performs better than structure based models (i.e., GCN and DeepWalk). The explanation is that time-series information (time and sequence information) is more suitable to reflect the law of popularity growth and accordingly contributes more to the high prediction performance. This conclusion is consistent with the information shown in literature~\citep{gao2014effective} that the temporal feature is the most effective feature type for popularity prediction.

\begin{table}[t]
\caption{Ablation study of each module in TCAN on both datasets.}
\begin{tabular*}{\tblwidth}{@{} LLLLLLL @{}}
\toprule
Dataset &\multicolumn{3}{L}{Weibo} &\multicolumn{3}{L}{APS}\\
\midrule
Metrics& MSLE& MAE& $R^2$ & MSLE& MAE& $R^2$ \\
\midrule
TCAN-G&2.687 &1.230 &0.473&1.384 &0.912 &0.407 \\
TCAN-S&2.356 &1.094 &0.538&1.315 &0.896 &0.437 \\
TCAN-NT&2.699 &1.169 &0.471&1.351 &0.902 &0.421 \\
TCAN-PL &2.221&1.082&0.564&1.249&0.862&0.465\\
\textbf{TCAN}&\textbf{2.007} &\textbf{1.032} &\textbf{0.607}&\textbf{1.201} &\textbf{0.845} &\textbf{0.486} \\
\bottomrule
\end{tabular*}
\label{t4}
\end{table}

\subsection{Ablation Study}
To illustrate the effectiveness of each component in TCAN, we build several variants of our model to conduct an ablation study. 
\begin{itemize}
\item \textit{TCAN-NT}: We simply remove the TE module while keeping others unchanged. It is used to verify the importance of time embedding in our model.
\item \textit{TCAN-PL}: This variant only employs the periodicity function $\phi_p(t)$ and linearity function $\phi_l(t)$ in TE. It is used to compare the effect of adding non-linearity scaling on the results.
\item \textit{TCAN-G}: In this variant, we keep the CGAT and TE modules while removing the CSAT module. 
\item \textit{TCAN-S}: In contrast to TCAN-G, we only use the TE and CSAT modules to capture the time-awareness cascade sequence information.
\end{itemize}

\begin{figure*}[b]
\centerline{\includegraphics[width=1\textwidth]{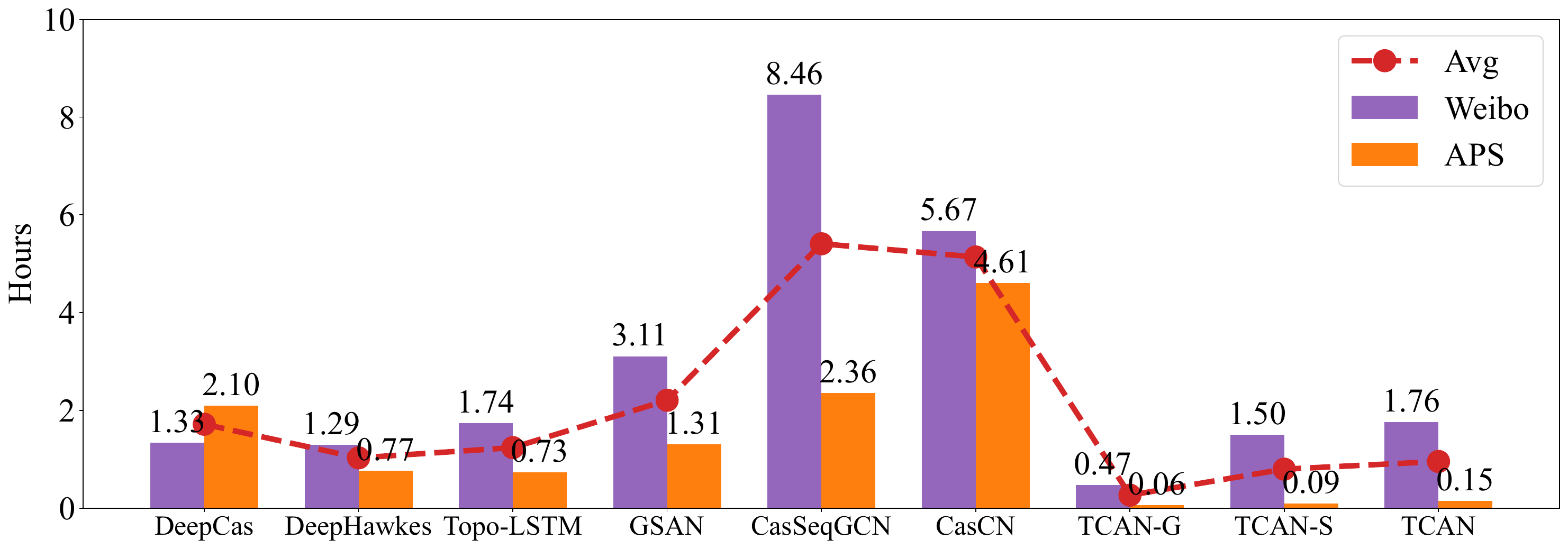}}
\caption{The running time cost (hours) of deep learning based methods on Weibo and APS datasets. Avg denotes the average time of two datasets.}
\label{fig10}
\end{figure*}

Table \ref{t4} shows the ablation study results. We  draw the following conclusions: 

\textbf{(1)} TCAN-S performs better than TCAN-G, and the reason behind the result is similar to the explanation mentioned in Section~\ref{ehs}-(3). Furthermore, both TCAN-S and TCAN-G are better than their corresponding enhanced vanilla models, i.e., TempRNN and TempGCN (in Table \ref{t3}). Therefore, our designed graph embedding module and sequence embedding module can indeed contribute to good performance. 

\textbf{(2)} Removing TE (TCAN-NT) leads to significant performance degradation. It indicates that TE endows TCAN the ability to capture cascade temporality to support accurate predictions. More conveniently, we do not need the complex operations of path sampling or episode division. 

\textbf{(3)} When we add the non-linear scaling term, the prediction accuracy of TCAN is further improved (compared to TCAN-PL). It shows that the non-linear scaling property is an effective temporal attribute for the popularity prediction task.

\subsection{Efficiency analysis}
In this subsection, we count the running time cost of TCAN, variants of TCAN (TCAN-G and TCAN-S), and other deep learning based baselines, and Fig.~\ref{fig10} shows the corresponding results. We analyze the running efficiency of TCAN from the following two perspectives:

(1) From the perspective of the composition of TCAN: Since TE has few parameters and low complexity, adding TE to TCAN brings almost no extra running time cost, so the TE module is not included in our statistics. CGAT and CSAT are the two modules that TCAN executes sequentially, and they dominate the running time consumption. As Fig.~\ref{fig10} shows, TCAN-G runs faster than TCAN-S, which benefits from its ability to support parallel computing well. Also the running time of TCAN is less than or equal to the sum of the running times of TCAN-G and TCAN-S on both datasets, demonstrating the rationality of TCAN architecture design.

(2) From the perspective of comparability: 
TCAN runs faster than GSAN, CasSeqGCN, and CasCN and is comparable to DeepCas, DeepHawkes, and Topo-LSTM on Weibo. In particular, TCAN achieves the best results on APS, and the average running time of TCAN on both datasets is superior to others. This demonstrates that TCAN has both superior prediction performance and high running efficiency.

\subsection{Interpretability Study}
\begin{figure}[tp]
    \centering
    \subfigure[Observation time window of Weibo]{
    \includegraphics[width=6cm,height=4.5cm]{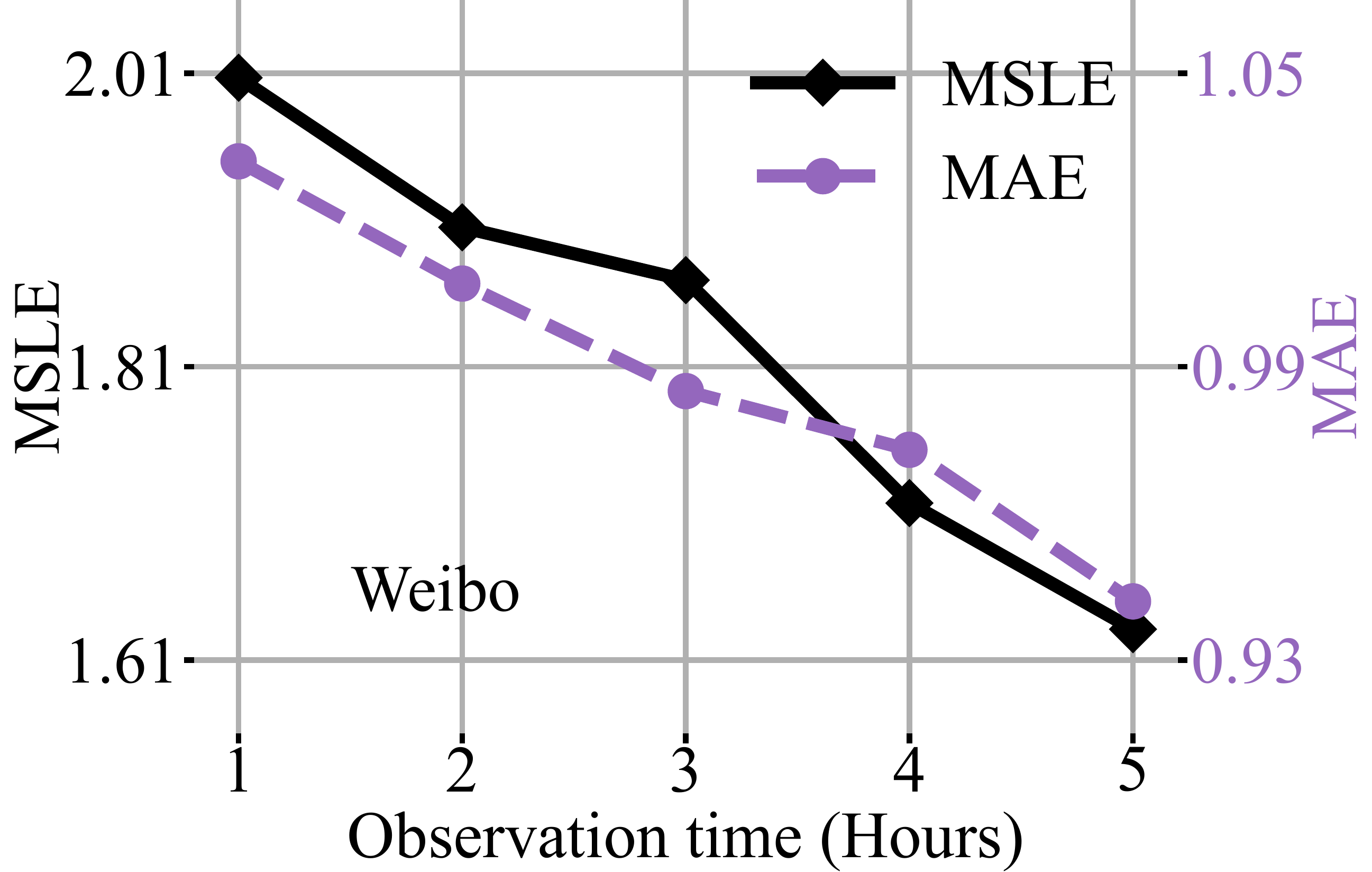}
    }
    \qquad
    \qquad
    \qquad
    \subfigure[Observation time window of APS]{
    \includegraphics[width=6cm,height=4.5cm]{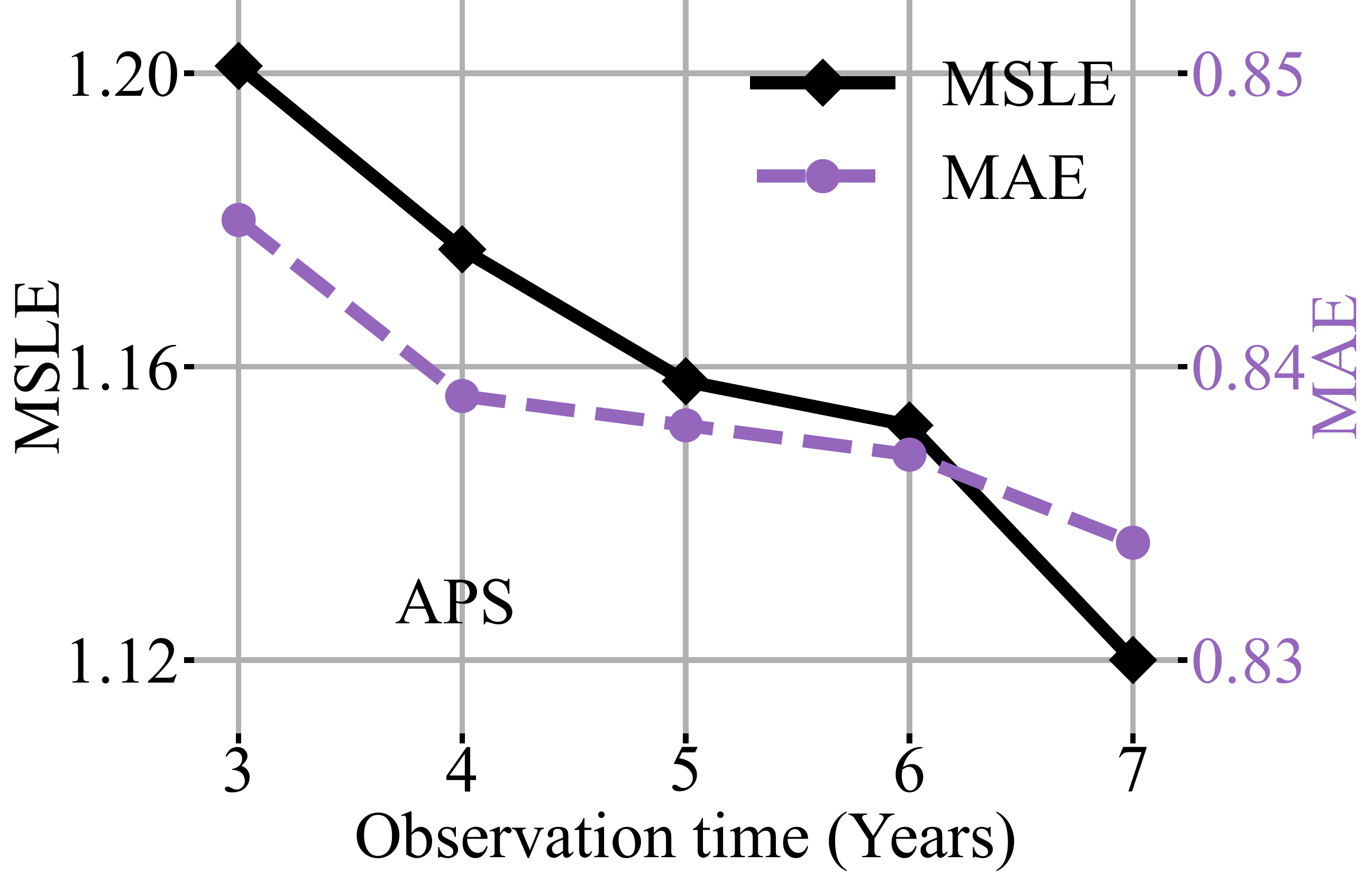}
    }
    \subfigure[Heatmap of Weibo]{
    \includegraphics[width=5.5cm,height=4.5cm]{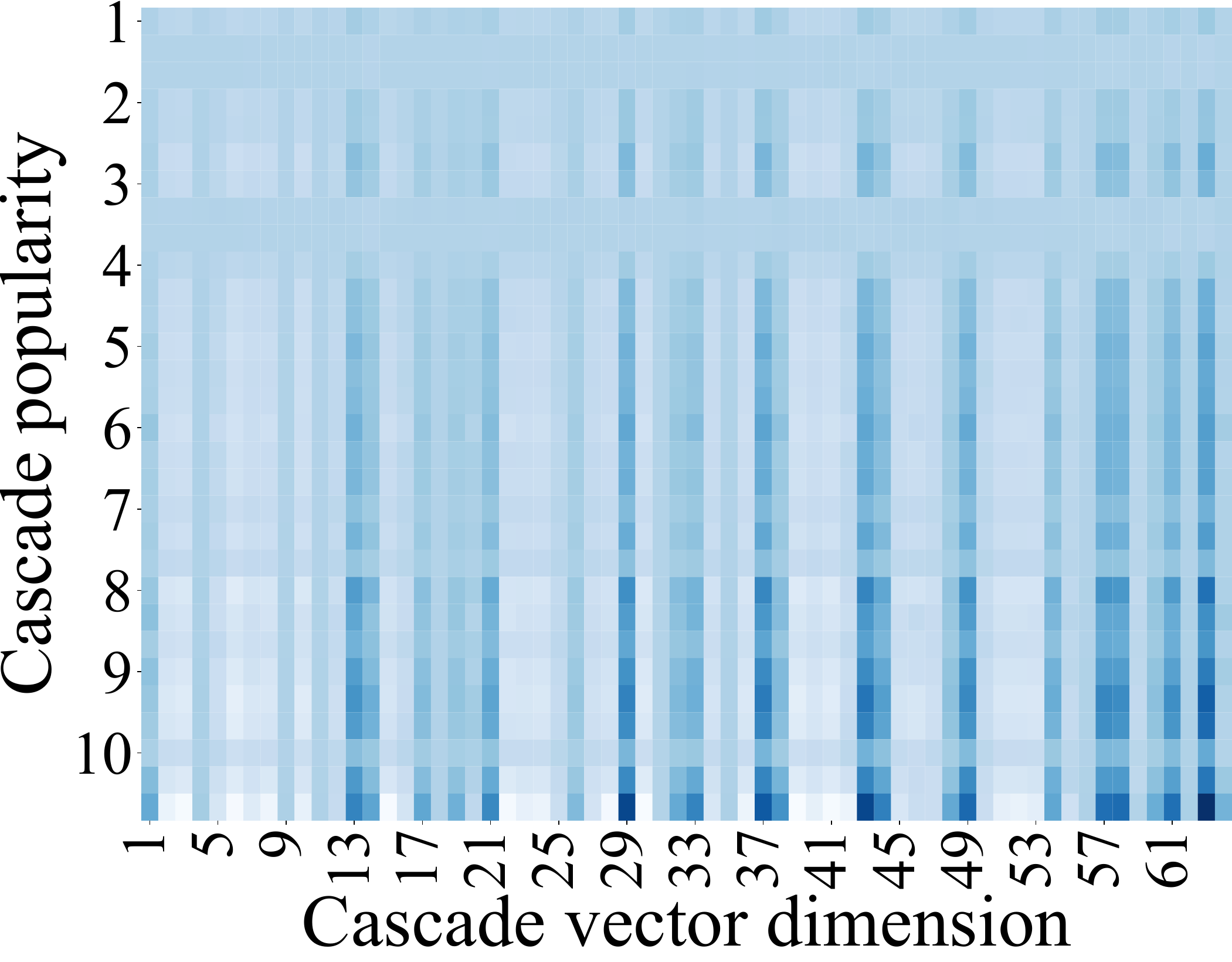}
    }
    \qquad
    \qquad
    \qquad
    \qquad
    \subfigure[Heatmap of APS]{
    \includegraphics[width=6cm,height=4.5cm]{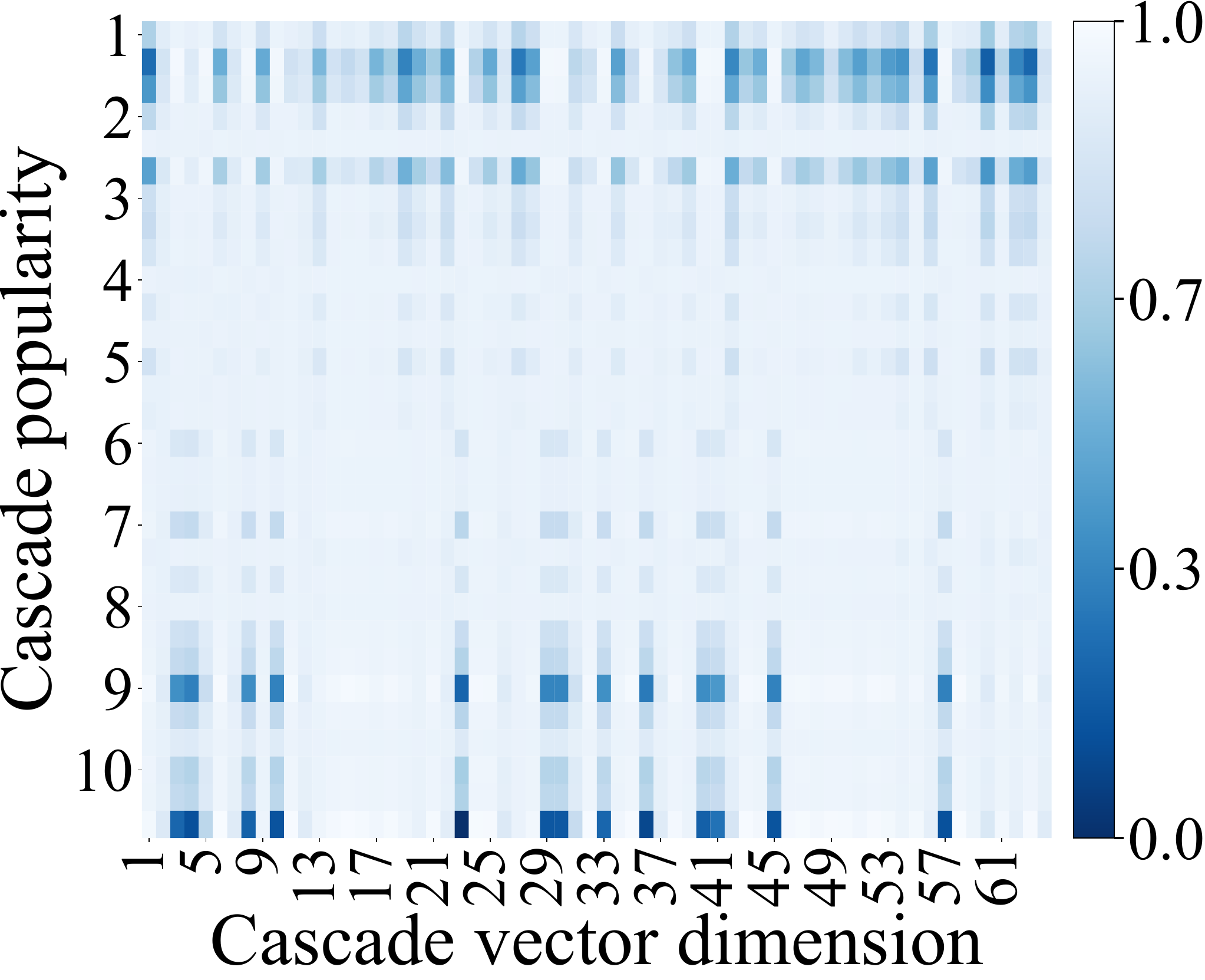}
    }
    \caption{Observation time impact and prediction visualization. (a) and (b) show the impact of observation time window size on Weibo and APS, where the double y-axis denotes two different metrics (i.e., MSLE and MAE). (c) and (d) show the prediction visual results on both datasets.}
\label{fig5}
\end{figure} 
    
\subsubsection{Observation time window analysis} Figure~\ref{fig5}(a) and (b) show the impact of different observation time window sizes on popularity prediction on both datasets. We can observe that MSLE and MAE show a downward trend along with in increase of the observation time window. That is because the cascade role information is getting richer as the time window size increases, which enables TCAN to learn more cascade graph structure and sequence features and make more accurate predictions of future popularity accordingly. 

\subsubsection{Prediction visualization} In order to find out how the prediction layer in TCAN can predict the popularity values of different sizes, we visualize the normalized cascade representation which generates from the second layer of the prediction layers. Figure~\ref{fig5}(c) and (d) show the visualization results, where the horizontal axis denotes the dimension ($x=1,...,64$) of the cascade representation, and the vertical axis denotes popularity scales($2^y, y=1,...,10$). For each popularity scale, we sample three cascades for visualization. We can observe that the cascade representations' visual pattern with different popularity scales is clearly differentiated. The cascade representation on the Weibo dataset presents a three-segment distribution according to the popularity scale (i.e., $y$ is in the interval [1,3], [4,7], and [8,10] respectively), and APS roughly presents four-segment distribution (i.e., $y$ is in the interval [1,2], [3,5], [6,7], and [8,10] respectively). This phenomenon shows that our model can learn a specific representation distribution for the cascade with different popularity scales to achieve accurate predictions.
\subsubsection{Some explanations about TE} 
\label{sete}
\begin{figure}[t]
    \centering
    \subfigure[The law of people's posting
behavior (Weibo)]{
    \includegraphics[width=6cm,height=4.5cm]{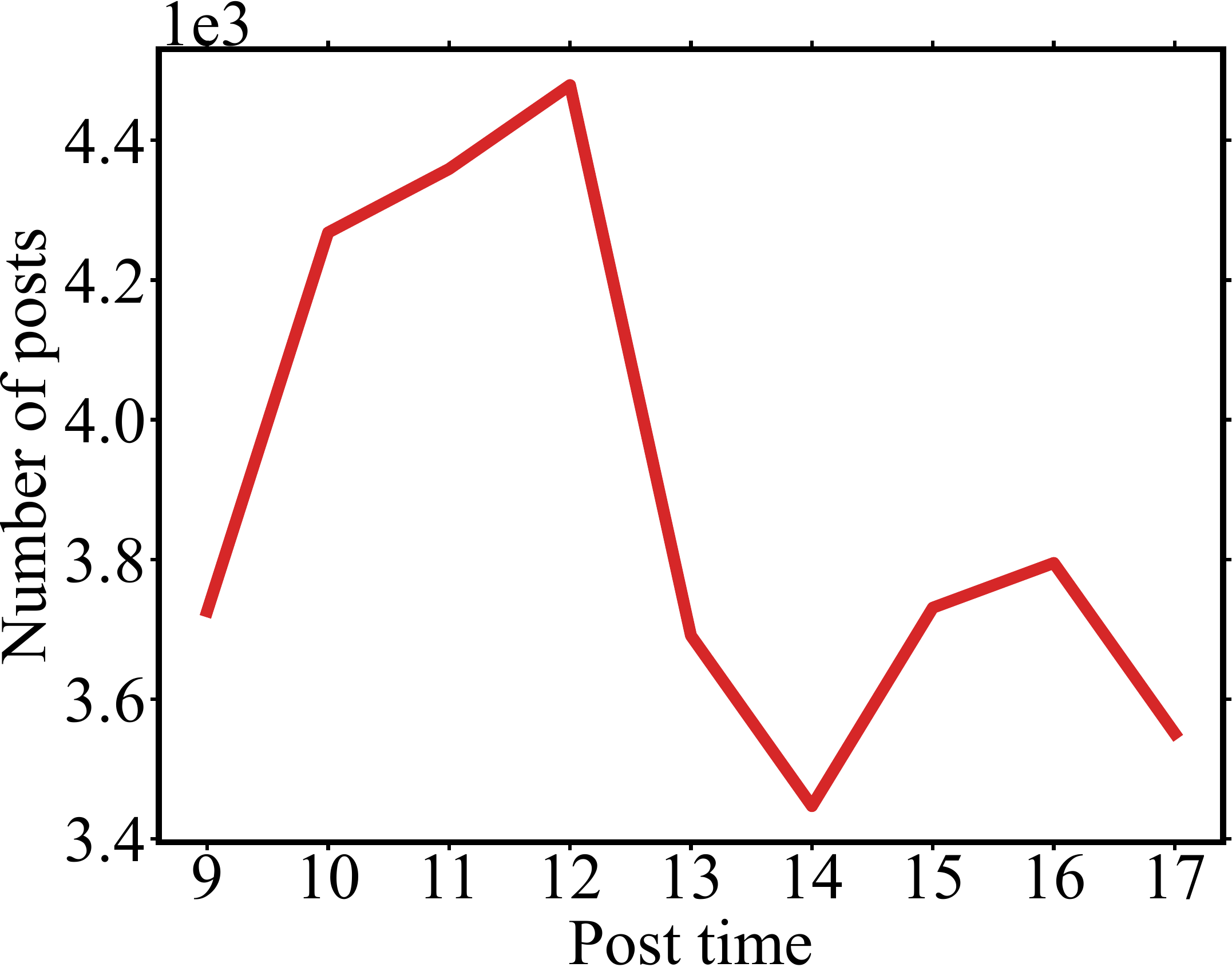}
    }
    \qquad
    \qquad
    \qquad
    \subfigure[The law of people's publishing
    behavior (APS)]{
    \includegraphics[width=6cm,height=4.5cm]{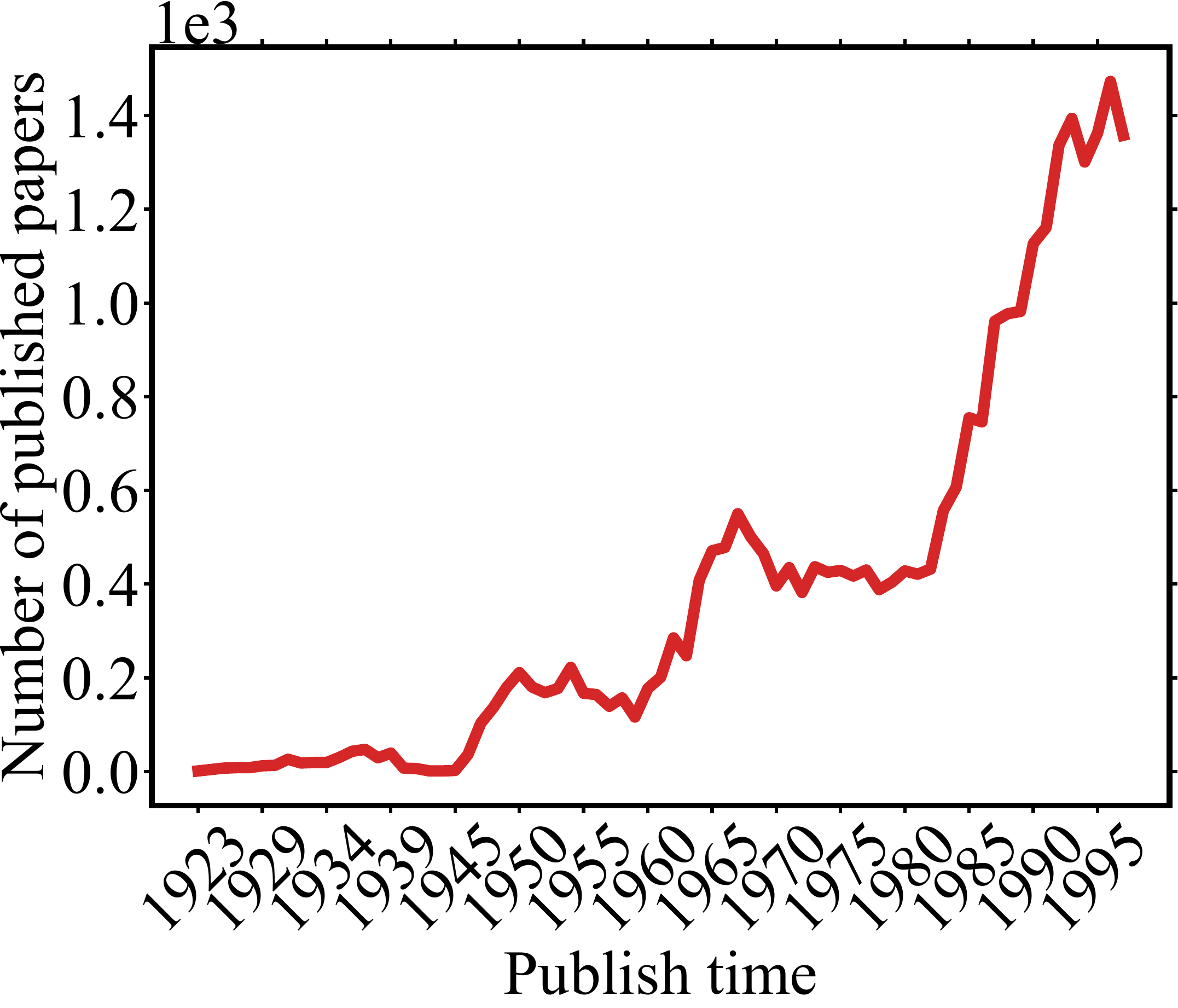}
    }
    \subfigure[The trend of popularity growth
    per hour (Weibo)]{
    \includegraphics[width=6cm,height=4.5cm]{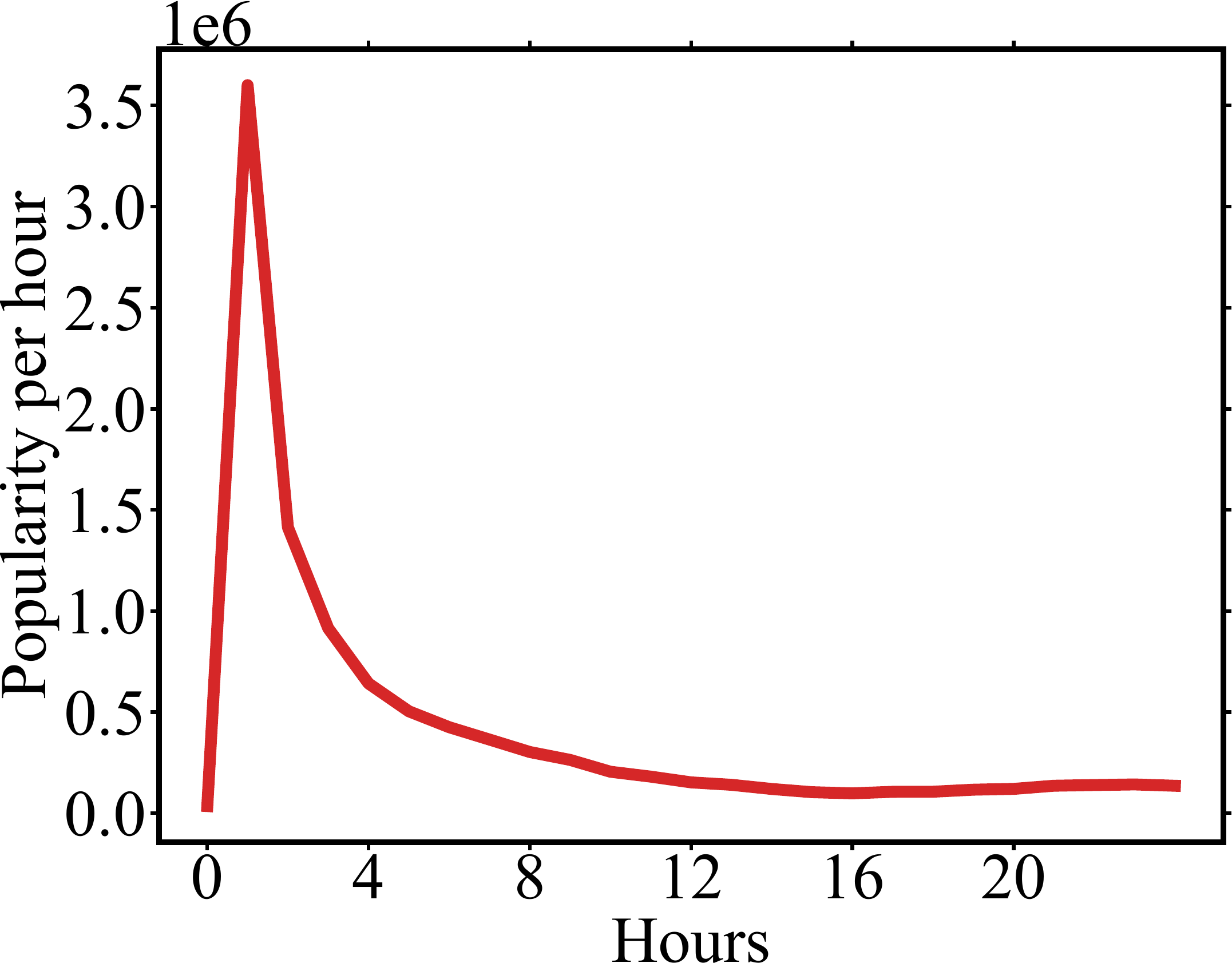}
    }
    \qquad
    \qquad
    \qquad
    \subfigure[The trend of popularity growth
    per year (APS)]{
    \includegraphics[width=6cm,height=4.5cm]{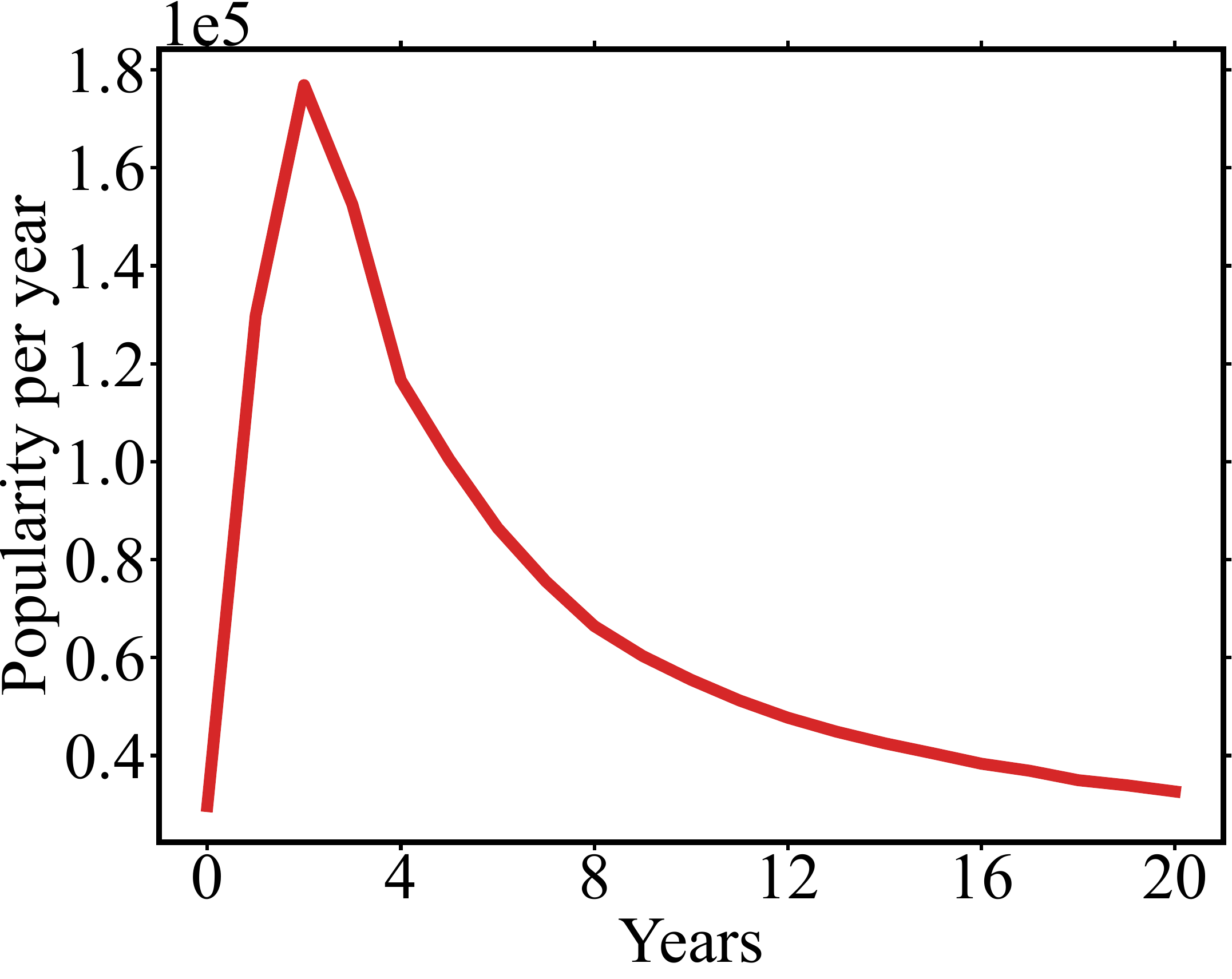}
    }
    \caption{Temporal attributes on both datasets. (a) and (b) show the number
    of the posted microblogs or papers changes with time. (c) and (d) show
    the trend of average popularity growth per hour or year.}
    \label{fig6}
    \end{figure} 
    
    \begin{figure}[t]
        \centering
        \subfigure[An example of a cascade graph]{
        \includegraphics[width=5.5cm, height=4.5cm]{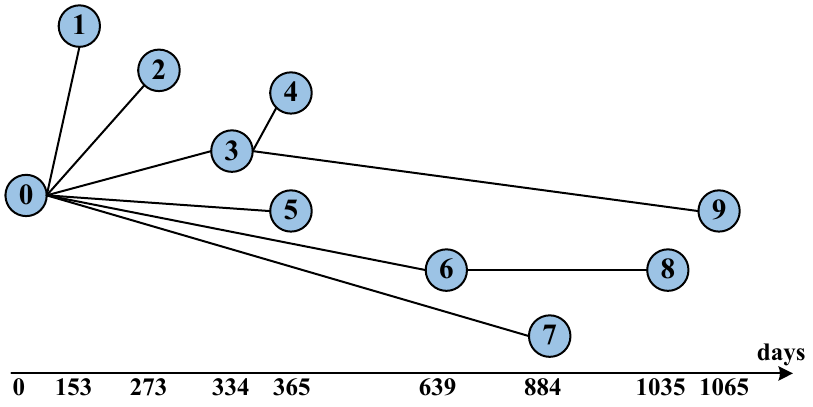}
        }
        \qquad
        \qquad
        \qquad
        \subfigure[The first layer]{
        \includegraphics[width=5.5cm,]{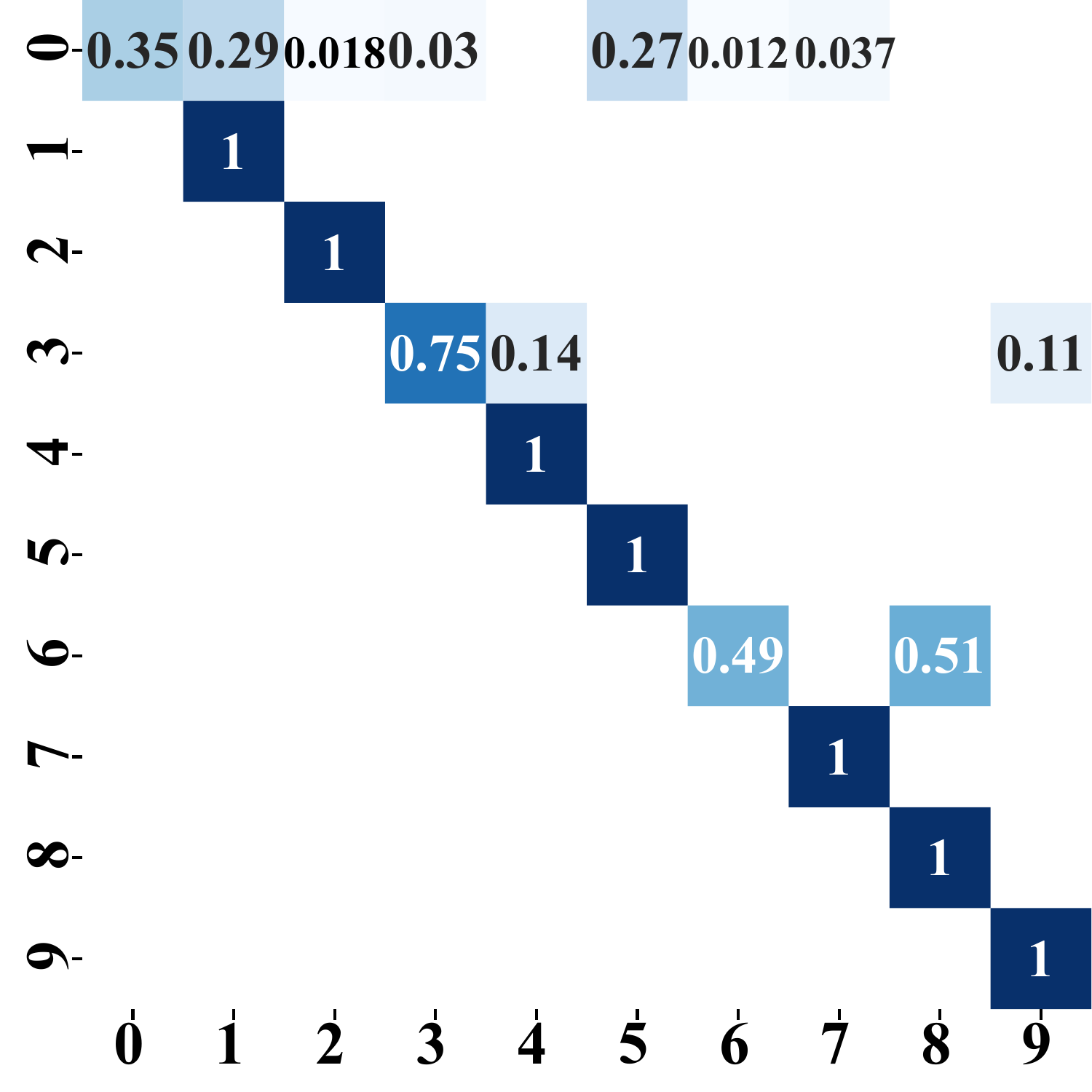}
        }
        \subfigure[The second layer]{
        \includegraphics[width=5.5cm]{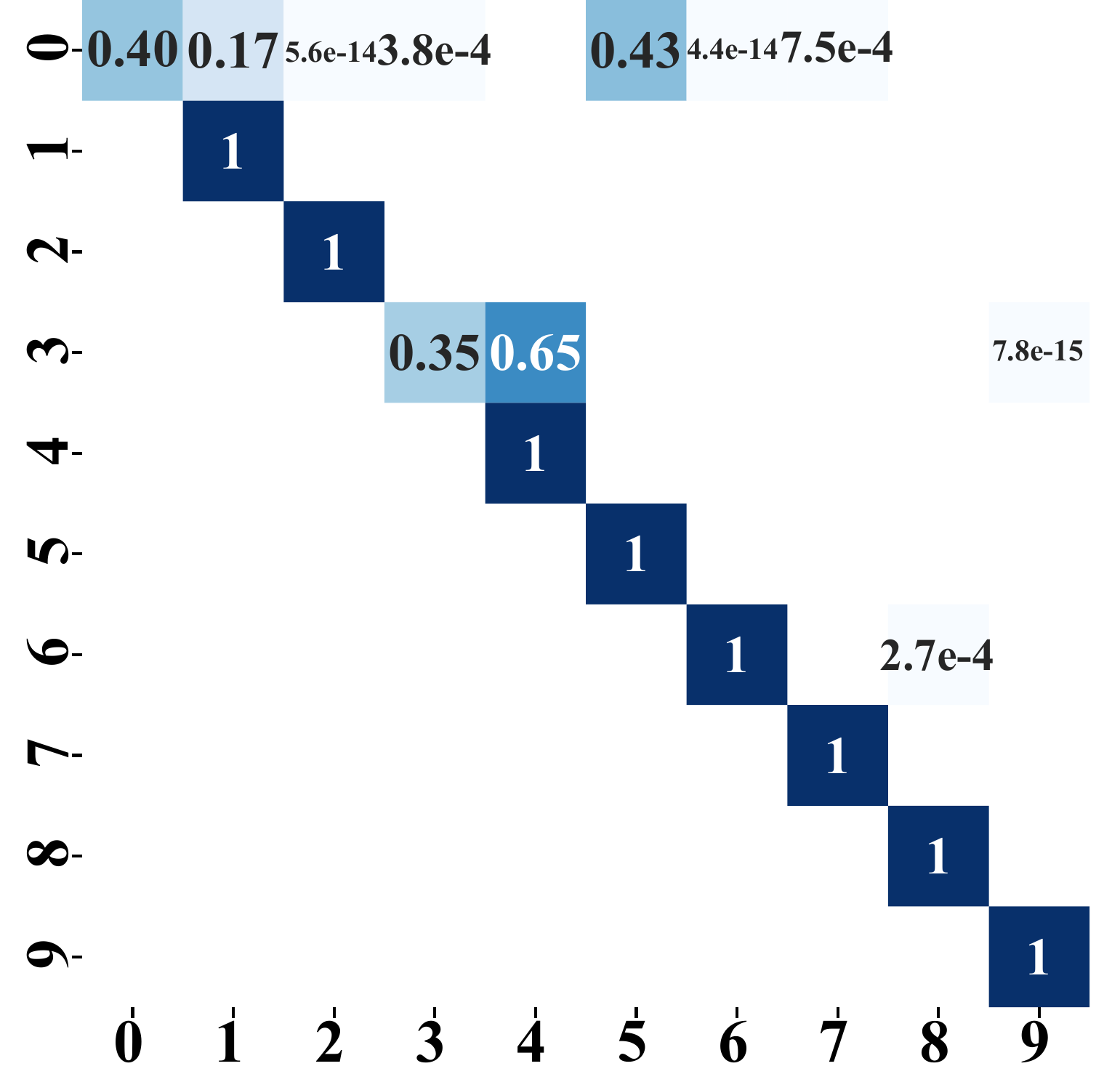}
        }
        \qquad
        \qquad
        \qquad
        \subfigure[The third layer]{
        \includegraphics[width=5.5cm]{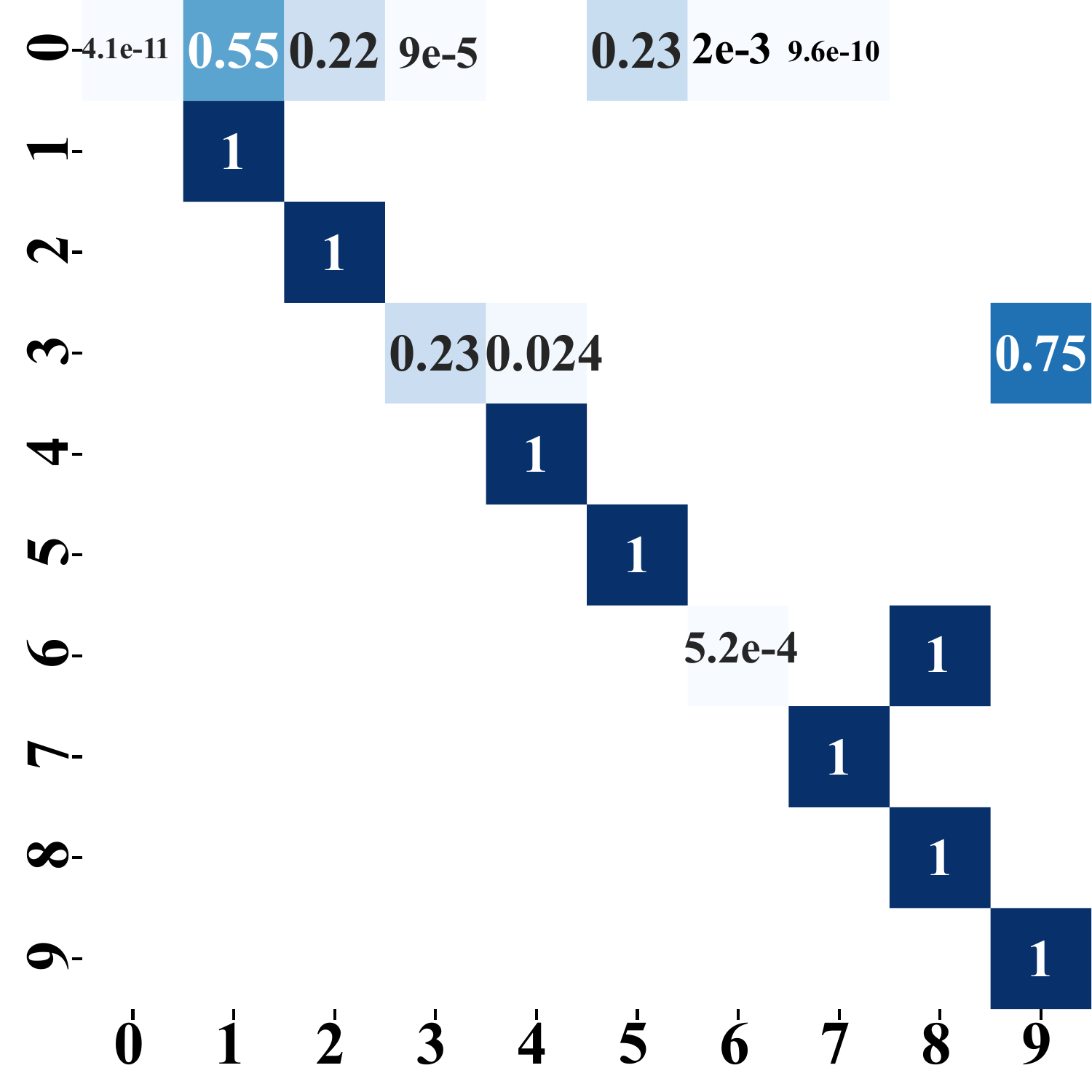}
        }
        \caption{Distribution of attention scores in CGAT. Figure (a) shows a cascade example from APS. Figure (b)-(d) show the distribution of its attention scores in different layers of CGAT.}
        \label{fig7}
        \end{figure} 
        
    \begin{figure}[tp]
    \centering
    \subfigure[Incremental popularity for Weibo]{
    \includegraphics[width=6cm,height=4.5cm]{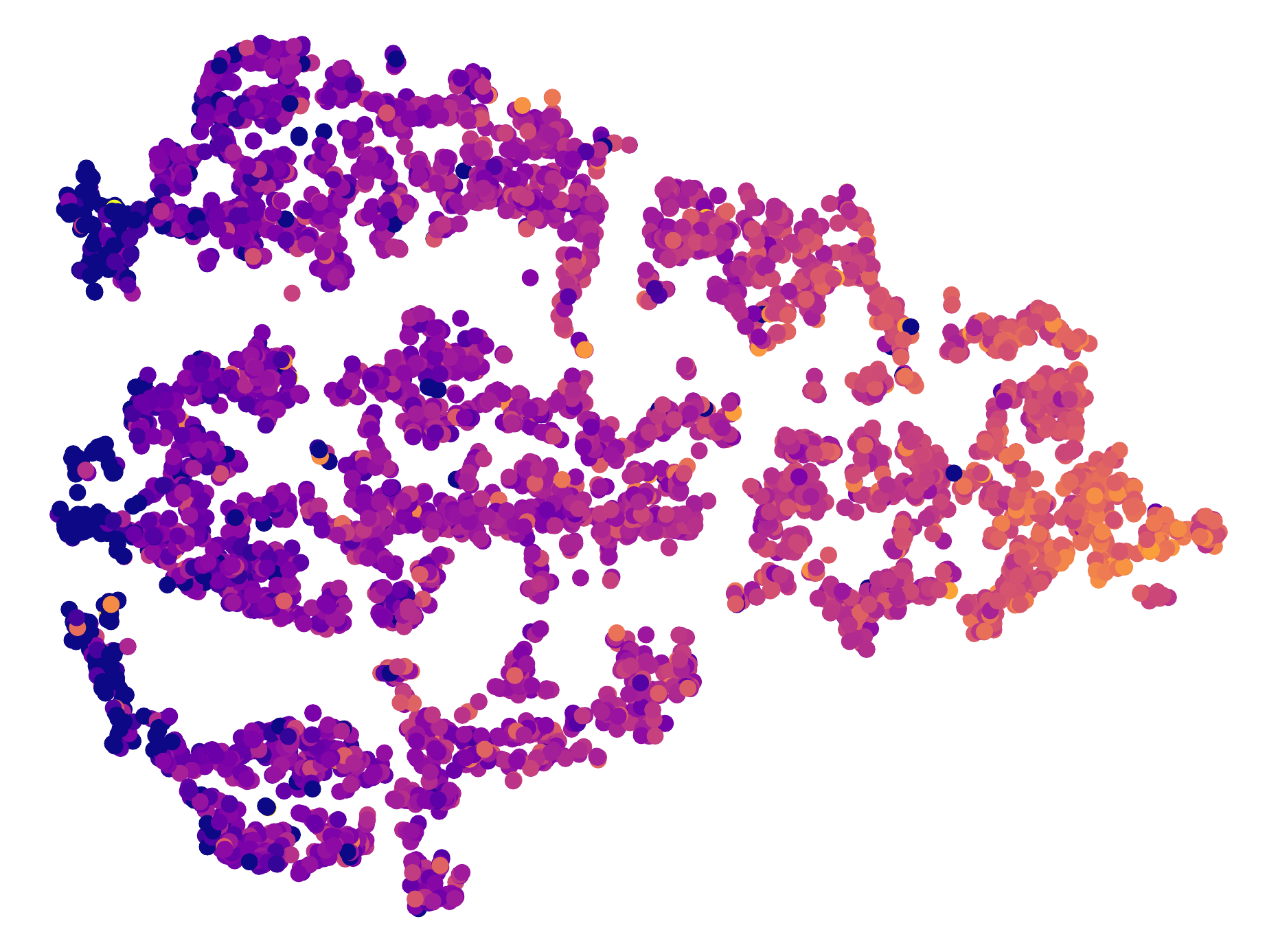}
    }
    \qquad
    \qquad
    \qquad
    \subfigure[Average repost time for Weibo]{
    \includegraphics[width=6cm,height=4.5cm]{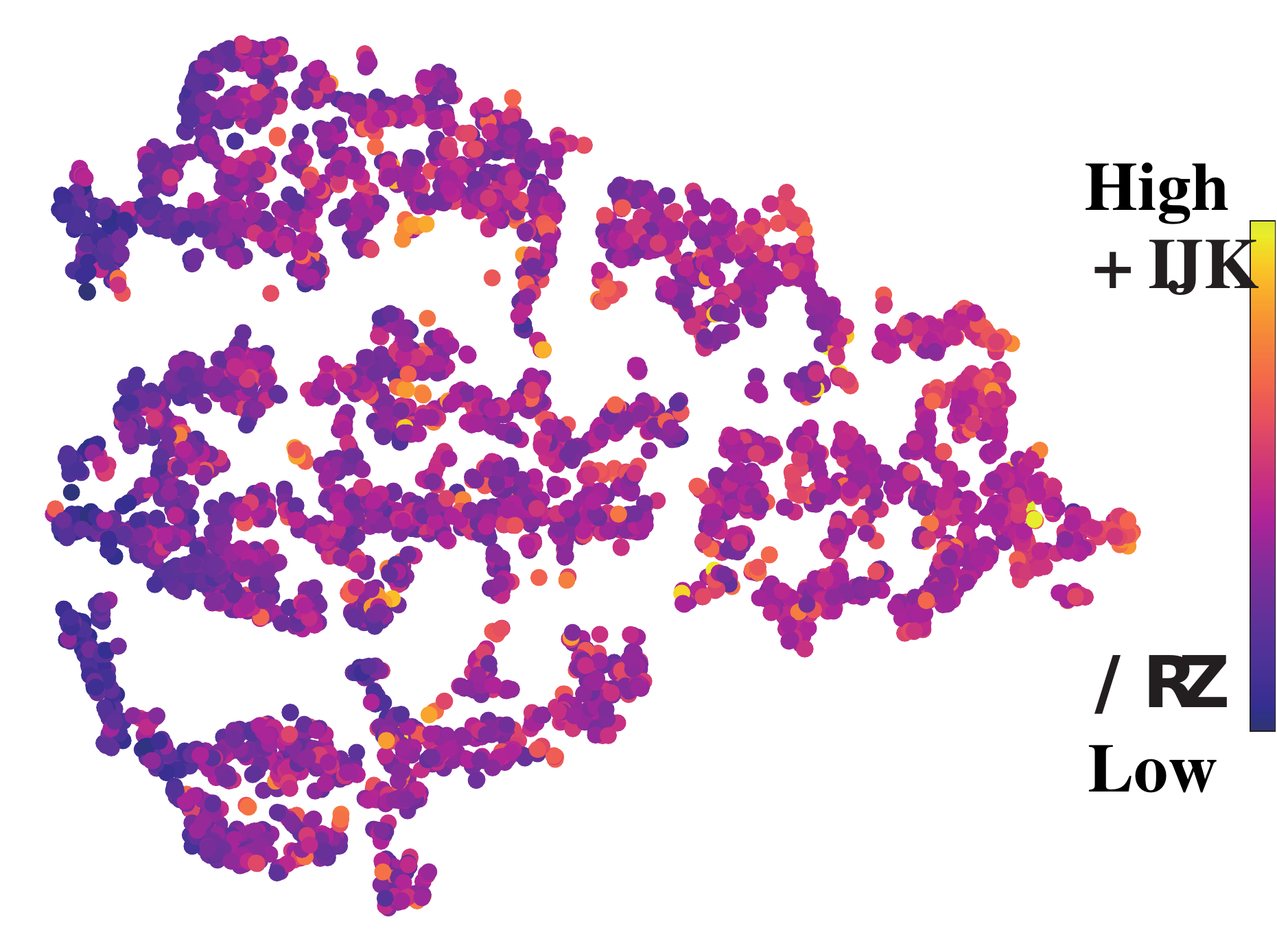}
    }
    \subfigure[Average SP length for Weibo]{
    \includegraphics[width=6cm,height=4.5cm]{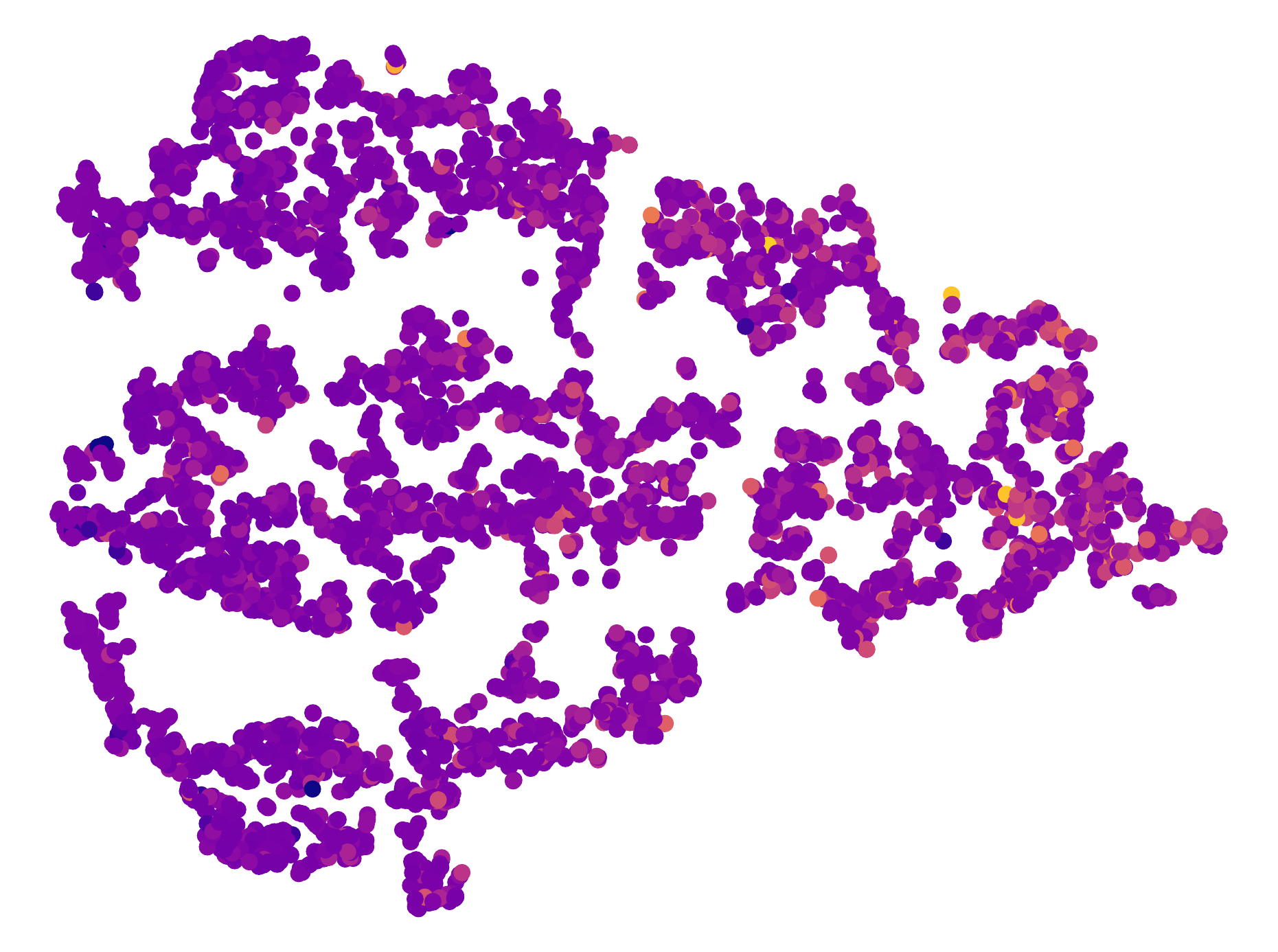}
    }
     \qquad
    \qquad
    \qquad
    \subfigure[Observed cascade size for Weibo]{
    \includegraphics[width=6cm,height=4.5cm]{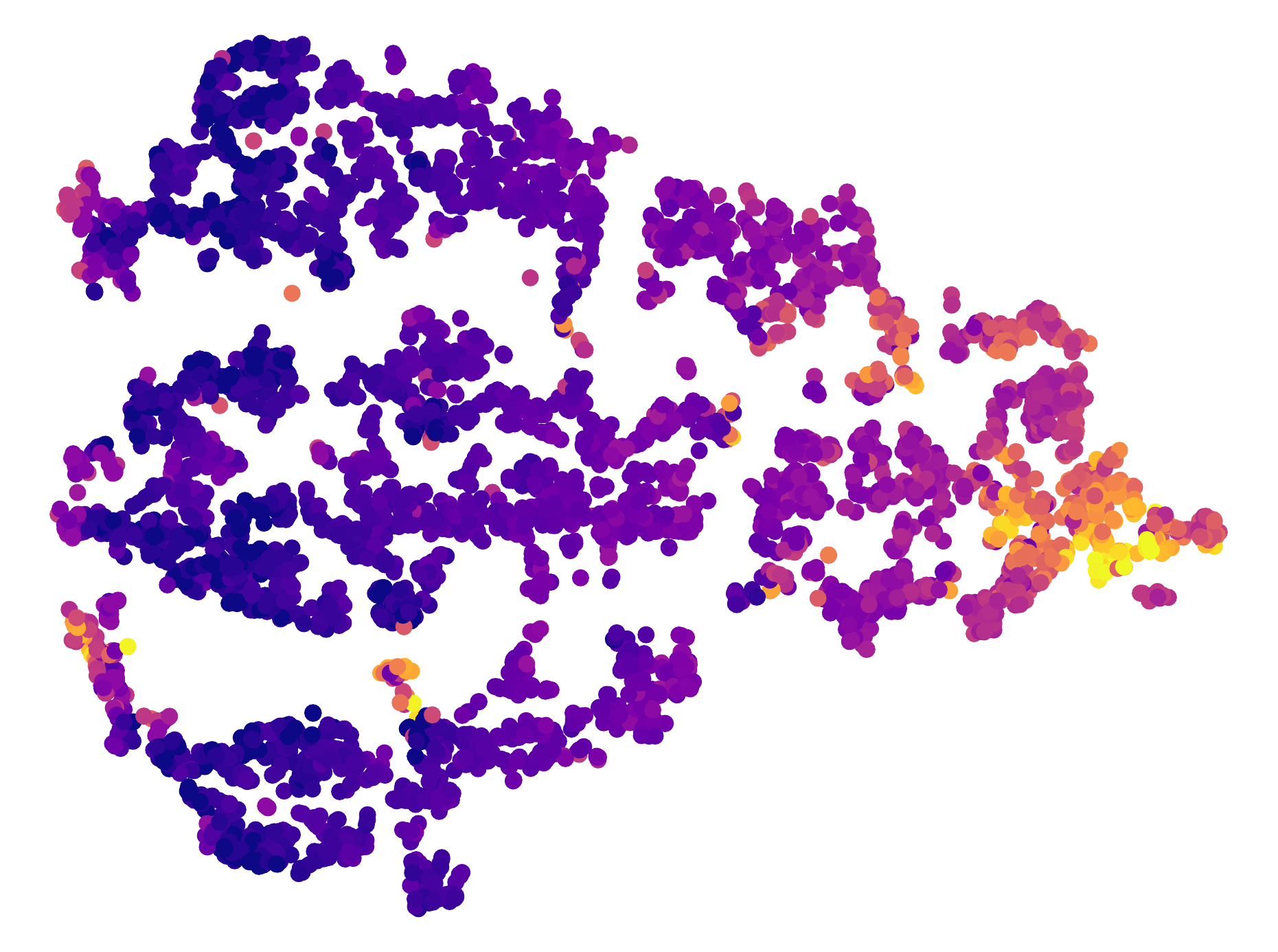}
    }
    \subfigure[Incremental popularity for APS]{
    \includegraphics[width=6cm,height=4.5cm]{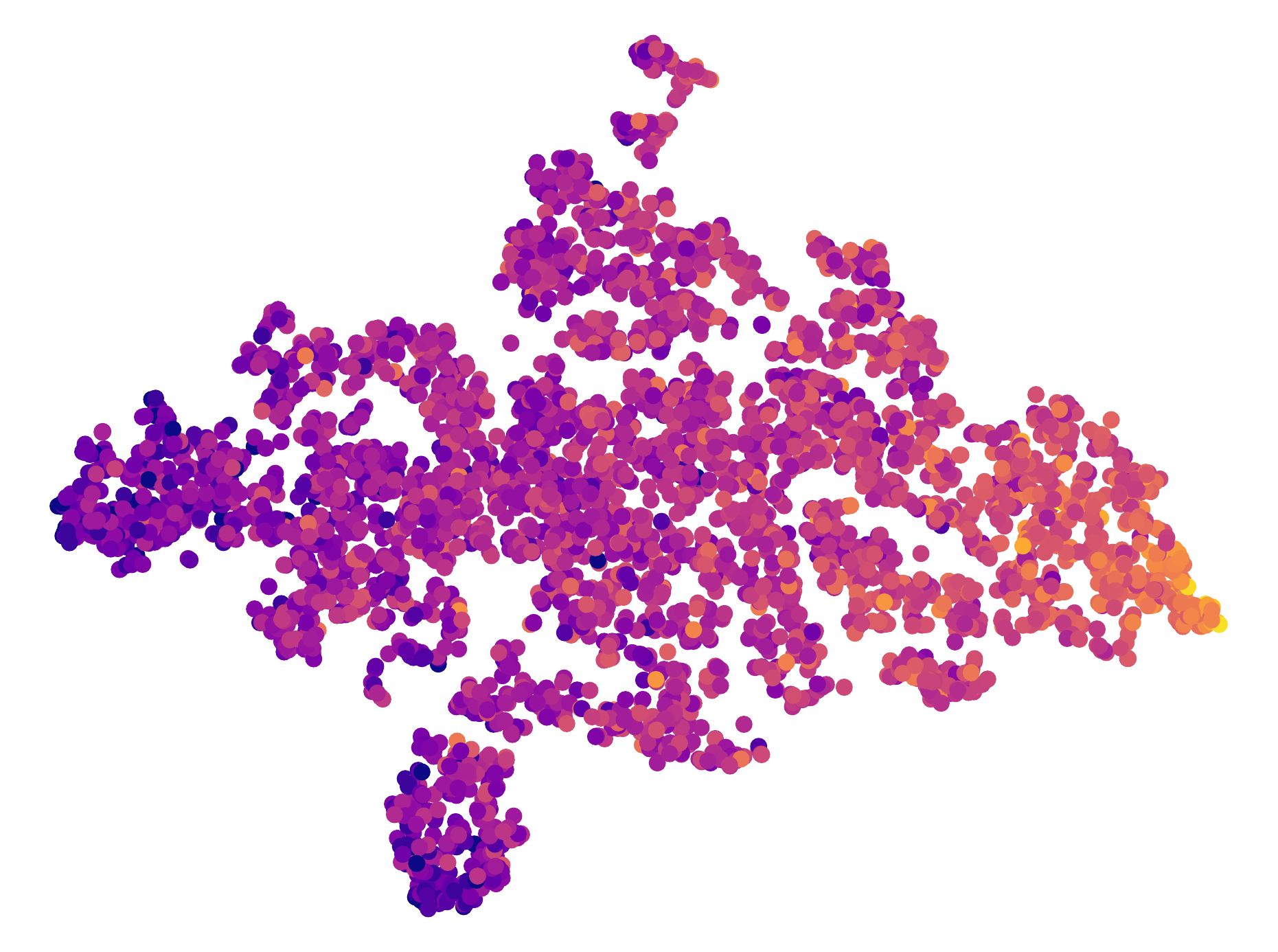}
    }
     \qquad
    \qquad
    \qquad
    \subfigure[Average repost time for APS]{
    \includegraphics[width=6cm,height=4.5cm]{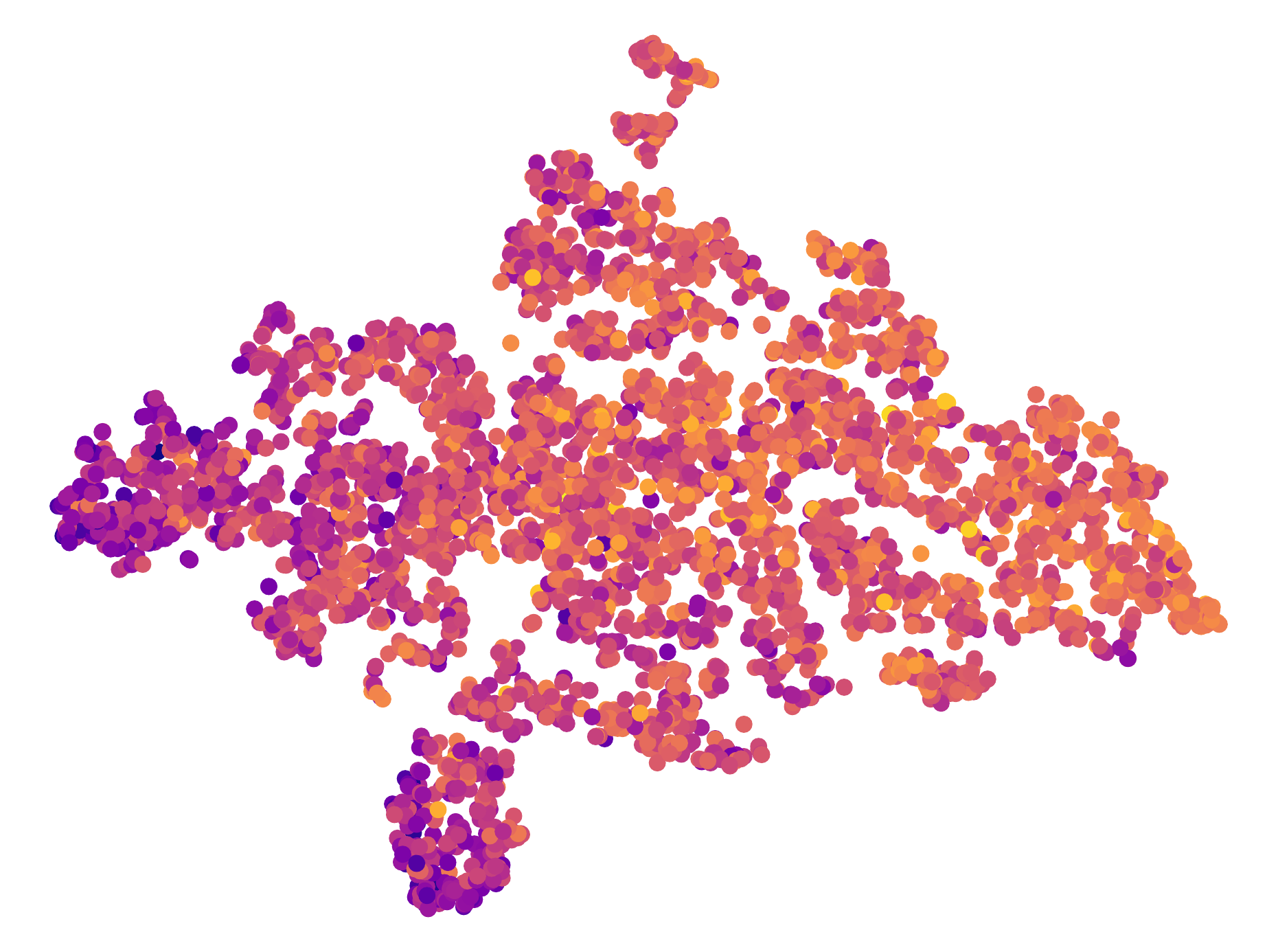}
    }
    \subfigure[Average SP length for APS]{
    \includegraphics[width=6cm,height=4.5cm]{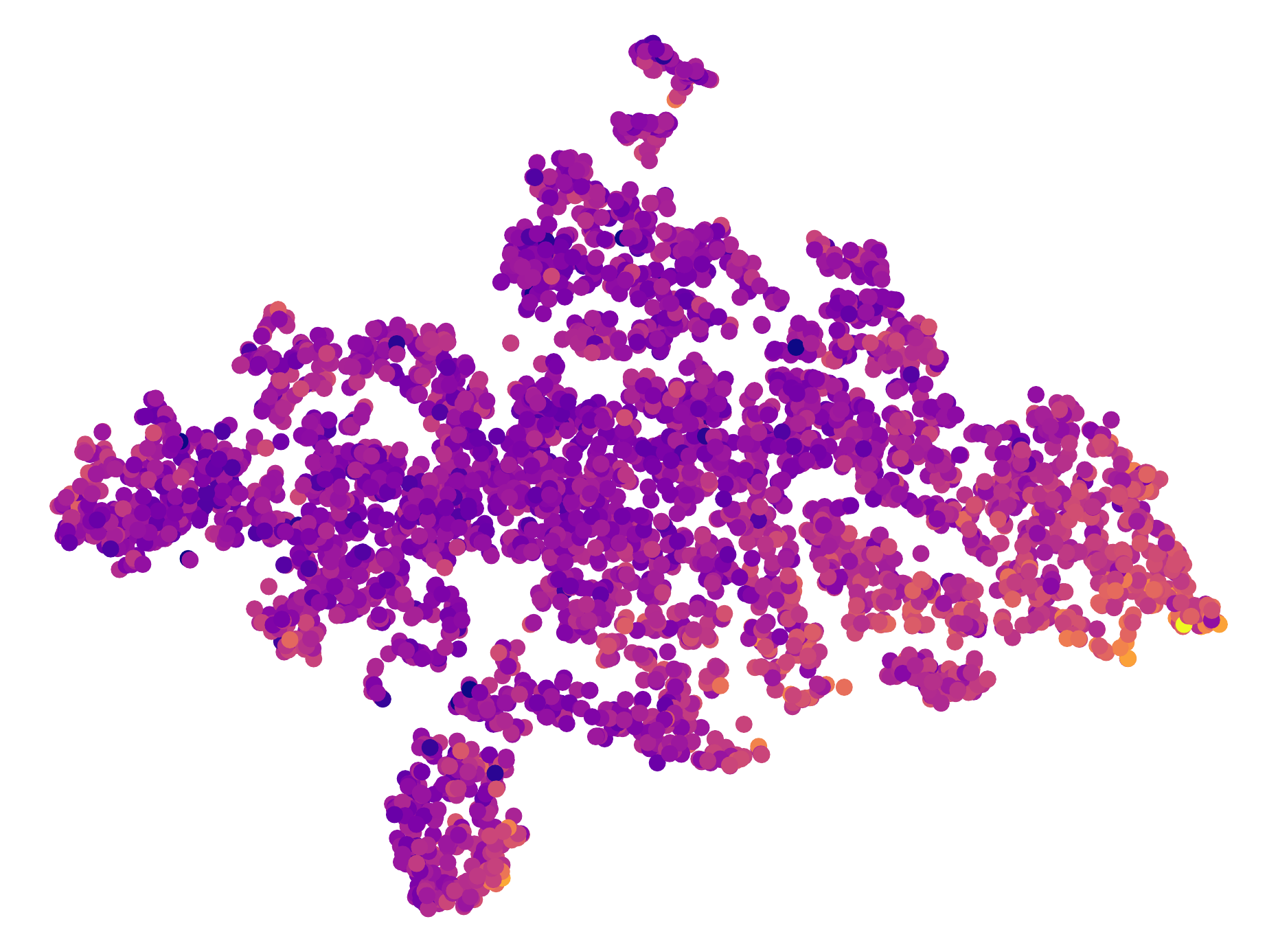}
    }
     \qquad
    \qquad
    \qquad
    \subfigure[Observed cascade size for APS]{
    \includegraphics[width=6cm,height=4.5cm]{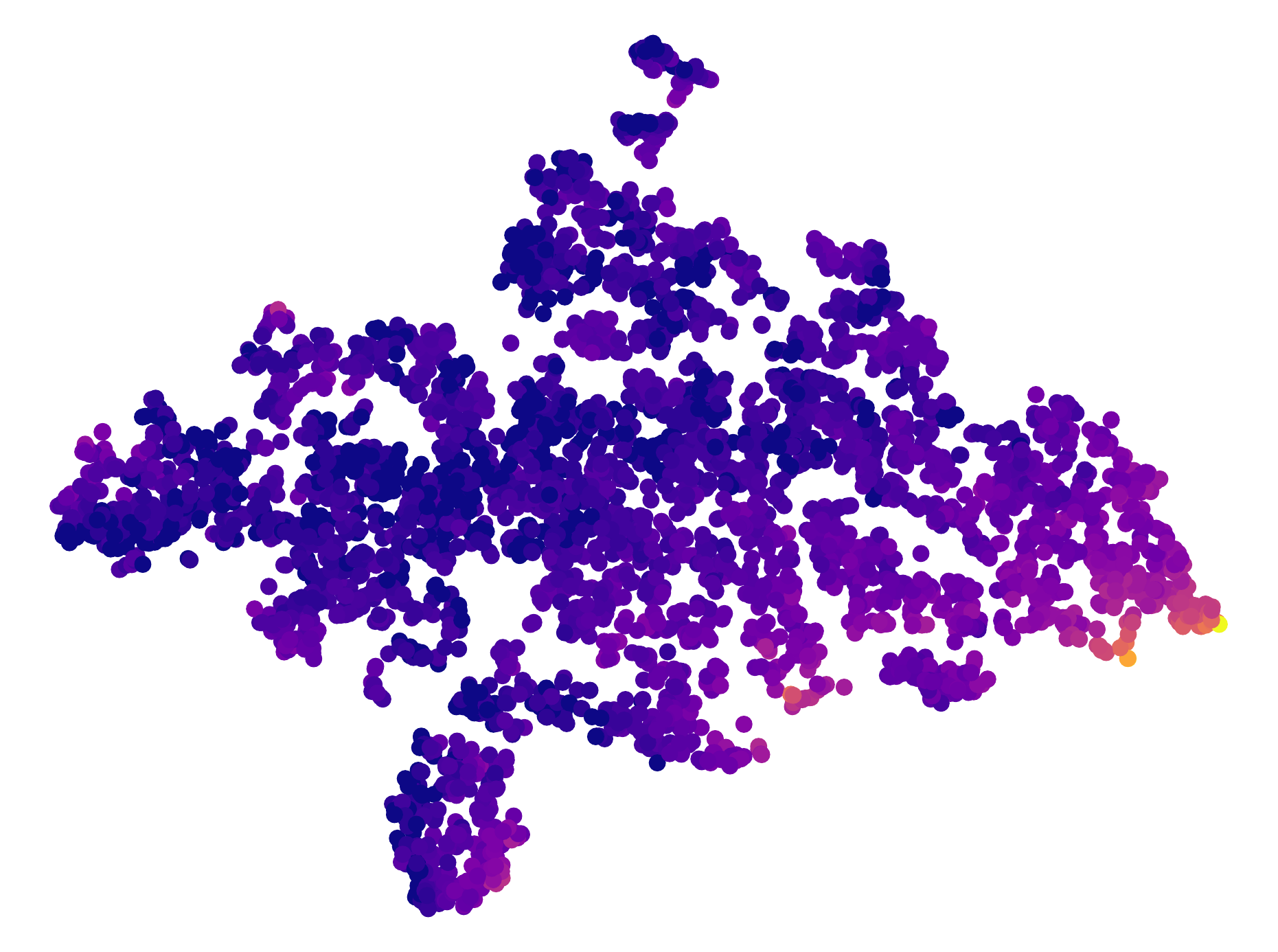}
    }
    \caption{The visualization of 2D representation distribution of TCAN, where the color represents the ground truth or cascade properties.
    }
    \label{fig8}
    \end{figure}  
We use statistical methods to show the temporal attributes (i.e., periodicity, linearity, and non-linear scaling)  captured by our TE correlate with the people's posting behavior and popularity growth. 

\textbf{(1)} Figure~\ref{fig6}(a) and (b) show the law of people's posting/publishing behaviors. 
In Weibo, people's posting behavior is influenced by circadian rhythms. More people post microblogs during the day, and there is a periodic fluctuation in Figure~\ref{fig6}(a). 
In APS, people's publishing behavior may be related to political and economic factors, and there are many notable peaks and troughs overall in Figure~\ref{fig6}(b), showing a cyclical upward trend with time. Observing in time intervals (e.g., [1949,1959] and [1990,1997]), the pattern of people's publishing behavior contains periodic ingredients.

\textbf{(2)} Figure~\ref{fig6}(c) and (d) tell us, the changing trends of popularity on different time scales (hour or year) are similar for the two datasets, i.e., the popularity per unit time increases linearly in the first short time and then decreases non-linearly over time. It is worth noting that we do not distinguish between positive and negative trends, they are considered as identical features.

We conclude from the above analysis that people's publishing behaviors and popularity changes reflect similar attributes to that of time. Therefore, from the practical level, our time embedding vector also can represent the user publishing behaviors and popularity changes to help us make better predictions.

\subsubsection{Some explanations about CGAT} 
We select a cascade from the APS dataset as an example (as shown in Figure~\ref{fig7}(a)) to analyze the attention scores collected from the trained CGAT. We draw the attention scores by the heat map to conveniently observe the attention distribution of each CGAT layer. Figure~\ref{fig7}(b)-(d) show the average of the attention scores of a total of four heads from three CGAT layers. We can observe two phenomena:

\textbf{(1)} In the first row of Figure~\ref{fig7}(b)-(d), the attention scores of leaf neighbors with short repost times are almost higher than that of non-leaf neighbors or leaves with long repost times. This phenomenon indicates that the root node is more inclined to aggregate the leaf nodes with a short repost time, which can improve the proportion of important one-hop leaf nodes in feature pooling. That is because the popularity prediction is a regression task, and one-hop leaf nodes are abundant in the cascade and contribute a lot to popularity (as shown in the cascade path length distribution in Figure~\ref{fig4}). Therefore, our CGAT can effectively aggregate the 
one-hop leaf nodes according to the repost time to obtain high-quality cascade graph representations. 

\textbf{(2)} In the first layer, the neighbors of a specific node (e.g., node 3 and 6) have similar attention scores, but the last two layers show an opposite distribution of attention scores, where nodes in the third layer pay more attention to the new influenced neighbors (e.g., node 9's score is larger than node 4's score). These phenomena indicate that our CGAT can exploit new high-hop nodes in the high-level layers to explore the possibilities of more nodes joining.

\subsubsection{Some explanations about TCAN}
We intuitively interpret the cascade representations learned by TCAN through visualization techniques. T-SNE is first employed to map the learned cascade representations to a 2D space on Weibo and APS test data, and then we draw these feature points on a scatter plot. As shown in Figure~\ref{fig8}, each point represents a cascade in test set, and the color refers to the value of incremental popularity or cascade properties (i.e., the average repost time, the average shortest path (SP) length, and the observed cascade size). The dark cascade points correspond to small feature values.

In Figure~\ref{fig8}(a) and (e), we see a clear color gradient on incremental popularity. This phenomenon indicates that the cascade representations learned by TCAN are distinguishable in terms of the popularity of different sizes, and thus TCAN can deal with the outbreak and non-outbreak cascades.

For the cascade properties, they color the learned representations by TCAN. If the color pattern of a cascade property is consistent with increment popularity, we will believe that the cascade property is not only related to future popularity but also has a connection with learned representations. We can see from Figure~\ref{fig8}(b), (c), (d), (f), (g), and (h) that they show a similar color gradient to incremental popularity, i.e., high repost time, long average SP length and observed cascade size correspond to large incremental popularity. Thus, these cascade properties are effective for popularity prediction. Furthermore, the average repost time, the average SP length, and the observed cascade size represent the temporal feature, structural feature, and sequence feature, respectively. TCAN can automatically learn meaningful cascade representations representing the above features for predicting future popularity, which provides a good explanation for TCAN's excellent performance. 

\section{Conclusion and Future Work} \label{sec6}
Finally, we summarized our work from theory and practice, and then we discussed some limitations of our work as well as several directions of our future research.

Theoretically, we proposed a deep learning framework (TCAN) capable of fully mining cascade information for popularity prediction. TCAN injected the explicit time features into cascade nodes by a general time embedding method (TE) and fully explored the cascade role information by employing a cascade graph attention network (CGAT) and a cascade sequence attention network (CSAT). 

The main differences from the existing deep learning methods for popularity prediction were as follows: (1) We used the time embedding (our proposed TE) to reflect the temporality of cascades, avoiding the complex operations of sequence sampling and cascade episode dividing. (2) We used deep learning models (our proposed CGAT and CSAT) to fully extract features from cascade graph and cascade sequence including potential information such as the temporal relationship between nodes in the graph and the temporal context of a node in the sequence. 

The comparison experiments (Section~\ref{cwb}) conducted on two real-world datasets show that TCAN was superior to all baselines. Furthermore, the enhancement study (Section~\ref{ehs}) demonstrated that TE was general and effective, and the interpretability study (Section~\ref{sete}) provided evidence for TE to capture the cascade temporality. The other visualization results indicated that TCAN can learn useful cascade representations for popularity predictions, which provided a good explanation for the high prediction performance.

Practically, TCAN can be extended to many application fields. For example, advertisement reposting prediction and rumor detection on social platforms, quantitative evaluation of literatures or authors on scientific citation networks, and epidemic prediction based on network data, etc.

{
In future work, we plan to exploit more general time embedding methods and effective cascading role learning architectures (e.g., powerful GNNs and Transformers) to further enhance the  popularity prediction performances. The research studies on cascade datasets are rare, and we will summarize the statistics and preprocessing methods of existing datasets, try to obtain high-quality, clean, and diverse cascade samples, and explore ways to generate synthetic datasets.}
Besides, there are plenty of studies working on the popularity prediction task, and most of them including our TCAN are devoted to predicting the final popularity or popularity growth before the end time, while multi-period popularity prediction is rarely studied. The multi-period popularity prediction is a more complicated time series forecasting problem that highly relies on mining cascade properties. This is a more meaningful yet more challenging task that can help decision-makers to make precise judgments in time.


\printcredits

\section*{Acknowledgments}
This work was partially supported by the National Natural Science Foundation of China (No. 61972272), the Natural Science Foundation of the Jiangsu Higher Education Institutions of China (No. 21KJA520008), Qinlan Project of Jiangsu Province of China, the National Science Foundation (Nos. 2038029, 1564097), Postgraduate Research \& Practice Innovation Program of Jiangsu Province (Nos. KYCX22\_3196, SJCX21\_1344), and an IBM faculty award.

\bibliographystyle{model5-names}

\bibliography{cas-refs}




\end{document}